\documentclass[aps,prb,showpacs,amsmath,amssymb]{revtex4}
\usepackage{graphics}
\begin{document}

\title{
  Single-particle excitations and phonon softening 
  in the one-dimensional spinless Holstein model  
}
\author{S. Sykora$^{a}$, A. H\"{u}bsch$^{a,b}$, K. W. Becker$^{a}$,
G.~Wellein$^{c}$, and H.~Fehske$^{d}$}

\affiliation{
  $^{a}$Institut f\"{u}r Theoretische Physik,
  Technische Universit\"{a}t Dresden, D-01062 Dresden, Germany} 
\affiliation{  $^{b}$Department of Physics, University of California, 
Davis, CA 95616, USA}
\affiliation{ $^{c}$Regionales Rechenzentrum Erlangen, Universit\"at 
  Erlangen-N\"urnberg, Germany}
\affiliation{ $^{d}$Institut f\"ur Physik, 
  Ernst-Moritz-Arndt Universit\"at Greifswald,
  D-17487 Greifswald, Germany}

\date{\today}

\begin{abstract}
We investigate the influence of the electron-phonon coupling in 
the one-dimensional spinless Holstein model at half-filling 
using both a recently 
developed projector-based renormalization method (PRM) and an refined exact
diagonalization technique in combination with the kernel polynomial 
method. At finite phonon frequencies the system shows a metal-insulator 
transition accompanied by the appearance of a Peierls distorted state 
at a finite critical electron-phonon coupling. We analyze the 
opening of a gap in terms of the (inverse) photoemission spectral 
functions  which are evaluated in both approaches.  Moreover, the PRM approach 
reveals the softening of a phonon at the Brillouin-zone boundary 
which can be understood as precursor effect of the gap formation.       
\end{abstract}


\maketitle

\section{Introduction}
Despite the many years of study of the electron-phonon interaction 
in metallic systems, there remain fundamental problems that yet have to be
resolved. Especially systems which suffer strong electron-phonon 
coupling in conjunction with strong electron-electron 
interaction are in the center of present interest. Examples are 
cuprate high-temperature superconductors\cite{Lanzara,Gunnarson},
colossal magnetoresistive manganites\cite{Millis}, 
or metallic alkaline-doped C$_{60}$-based compounds. 
Furthermore, in a wide range of quasi-one-dimensional
materials, such as MX chains, conjugated polymers or 
organic charge transfer complexes\cite{mat}, the itinerancy
of the electrons strongly competes with the electron-phonon
coupling which tends to establish e.g. charge-density-wave
structures. Then, in particular at half-filling, Peierls insulating 
phases may be energetically favored over the metallic state.
Many interesting questions arise not only with a view to the 
associated metal to insulator transition but also concerning
the form of the single-particle excitation spectra well below and
above the transition. At present,
there is a clear need of reliable theoretical methods to tackle 
these problems in terms of minimal microscopic models.

In this paper we study as the perhaps simplest realization of 
a strongly coupled electron-phonon system, the so-called 
spinless Holstein model, 
describing the local interaction between dispersionless 
longitudinal optical phonons and the density of the electrons at a 
given lattice site 
$n_i = c_{i}^\dagger c_{i}$: 
\begin{eqnarray}
\label{1} 
{\cal H} &=& 
- t \sum_{\langle i,j\rangle} 
( c_{i}^\dagger c_{j} + {\rm h.c.} )
+ \omega_0 \sum_i  \; b_i^\dagger b_i
+ g \sum_i \; (b_i^\dagger + b_i)n_i .
\end{eqnarray}
Here, $c^{\dagger}_{i}$ ($b^{\dagger}_{i}$) denote
fermionic (bosonic) creation operators of electrons (phonons), 
$i$ is the Wannier site index. The electron-phonon coupling constant and 
frequency of the Einstein mode is given by  $g$ and $\omega_0$, respectively.
Of physical concern for most applications are relatively small 
values of the adiabaticity ratio $\alpha=\omega_0/t\ll 1$, even though 
the anti-adiabatic limit $\alpha\gg 1$ is also a useful point of 
reference for an overall understanding of the physics of the 
Holstein model.

It is well-known that the one-dimensional Holstein model of
spinless fermions at half-filling has a quantum phase transition
from a Luttinger liquid (metallic phase) to an insulating phase 
with charge-density wave long-range order \cite{HF83,BGL95,FWHWBB04}.  
In the past, a large number of different analytical and numerical
methods have been applied to the Holstein model, 
in particular to determine the phase boundary 
between metallic and insulating behavior for the half-filled
band case. Much of the work is restricted to the 
one-dimensional case.  Mainly ground state 
properties were investigated
by means of strong-coupling expansions\cite{HF83}, variational\cite{ZFA89} 
and renormalization group\cite{CB84,BGL95} approaches, as well as
world-line 
quantum Monte Carlo\cite{HF83} and Green's-function Monte Carlo\cite{MHM96} 
simulations. More recently, exact diagonalization\cite{WF98} (ED) 
and density matrix renormalization group\cite{FWHWBB04,BMH98,FWH04,jecki}
techniques  were applied. The metal-insulator transition is accompanied by 
the appearance of a gap in the electronic spectrum which, however, 
can best be observed in the ${\bf k}$-dependent 
one-particle spectral functions. In a recent dynamical mean 
field treatment\cite{MeyerBulla} in conjunction with a numerical
renormalization group   approach it was suggested 
that the opening of the electronic gap is 
accompanied by the appearance of a low-energy phonon peak in the 
total phononic spectral function.  In fact, we shall show in this paper
that the phonon modes at the Brillouin-zone boundary become soft.
This  can be understood as a precursor effect of a  lattice 
instability leading to a Peierls state for electron-phonon coupling
strength $g$ larger than some finite critical  value $g_c$. Moreover,
we shall evaluate electronic one-particle spectral functions 
which should show the opening of the gap at the quantum phase transition.
The evaluation of Luttinger parameters is beyond
the scope of interest of this paper.

One of the main aims of the paper is to show that a newly developed  
projector-based renormalization method (PRM) for many-particle 
Hamiltonians \cite{Becker} can be applied to the spinless 
Holstein model, though an extension to the case with spin 
would also be possible.
In principle this method is applicable to infinitely large systems, to the 
whole parameter regime of the electron-phonon 
coupling   $g/\omega_0$, to any finite filling,  and to any
finite spatial dimension. In the present paper we restrict ourselves to 
the case of one dimension and to half-filling in order to 
study the transition from the metallic to the insulating 
phase. It will turn out the Peierls instability 
associated with the metal-insulator transition can very well be 
described in the present formalism.  
Since the application of the PRM, however, 
is accompanied by some approximations, the reliability of the
technique  will be tested by comparing its 
results with unbiased data from exact diagonalization (ED). 
In particular, the ${\bf k}$-dependent 
one-particle spectral functions from (inverse) photoemission will be 
analyzed within both approaches. As is seen, the formation of a 
charge gap  at some critical value $g_c$ of the electron-phonon coupling 
is related to the softening of phonon modes at the 
Brillouin-zone boundary, at least in the adiabatic regime. 
From the ED data we get additional valuable insights into the
behavior of the wave-vector resolved spectral function.
Moreover, a detailed characterization  of the  ground state,
e.g.~by the phonon distribution is possible.
We note that the present approach is not 
restricted to large dimensions as is the  
case of a recent DMFT approach \cite{MeyerBulla}. 

The paper is organized as follows. In Sect.~II we outline the 
PRM approach \cite{Becker}. The renormalization equations for 
the model parameters are derived non-perturbatively
and general expressions for the one-particle spectral functions
are discussed. In Sect.~III we discuss our exact 
diagonalization technique for the calculation of single-particle 
spectral functions of coupled electron-phonon systems. 
Sect.~IV presents the findings obtained within both approaches
{\ bf for the (I)PE spectra and phonon renormalization, in 
particular with respect} to the metal to Peierls insulator transition  
in the 1D Holstein model. The main results will be summarized in Sect.~V. 

\section{Renormalization of the Holstein model}

\subsection{Projector-based renormalization method}

The PRM \cite{Becker} starts from a decomposition of a given 
many-particle Hamiltonian ${\cal H}$ into an unperturbed 
part ${\cal H}_0$ and a 
perturbation ${\cal H}_1$
\begin{eqnarray}
\label{2}
{\cal H} &=&{\cal H}_0 + {\cal H}_1\,,
\end{eqnarray}
where we assume that the eigenvalue problem of ${\cal H}_0$ is solved 
\begin{eqnarray}
\label{3}
{\cal H}_0 \left| n^{(0)}\right\rangle &=& 
E_n^{(0)} \left| n^{(0)}\right\rangle\,.
\end{eqnarray}
The decomposition of ${\cal H}$ into ${\cal H}_0$ 
and  ${\cal H}_1$ should be done in such a way that 
${\cal H}_1$ contains no part which commutes with ${\cal H}_0$. 
Thus, ${\cal H}_1$ gives rise to transitions between 
eigenstates of ${\cal H}_0$ with different eigenenergies.
The presence of ${\cal H}_1$  usually
prevents the exact solution of the eigenvalue problem 
of the full Hamiltonian. Let us define a projection operator 
${\bf P}_\lambda$ by
\begin{eqnarray}
\label{4}
{\bf P}_\lambda {\cal A} &=& 
\sum_{m,n} \left| n^{(0)} \right\rangle \left\langle m^{(0)}\right|
\left\langle n^{(0)}\right| {\cal A} \left| m^{(0)}\right\rangle\,
\Theta(\lambda -|E_n^{(0)}-E_m^{(0)}|).
\end{eqnarray}
${\bf P}_\lambda$ is a super-operator which acts on ordinary
operators ${\cal A}$ of the unitary space. It projects on those parts of
${\cal A}$ which are formed by transition operators 
$\left| n^{(0)}\right\rangle \left\langle m^{(0)} \right|$ with energy 
differences $\left| E_n^{(0)} - E_m^{(0)} \right|$ less than a given 
cutoff $\lambda$, where $\lambda$ is smaller than the cutoff $\Lambda$ of the 
original model. Note that in Eq.~\eqref{4} neither 
$\left| n^{(0)}\right\rangle$ nor $\left| m^{(0)} \right\rangle$ 
have to be low-energy eigenstates of ${\cal H}_0$. However, their 
energy difference has to be restricted to values $\le \lambda$. 
Furthermore we define the projector 
\begin{eqnarray}
\label{5}
{\bf Q}_\lambda &=& {\bf 1} - {\bf P}_\lambda
\end{eqnarray}
on the high-energy transitions larger than the 
cutoff $\lambda$, (${\bf Q}_\lambda
{\bf P}_\lambda=0$). 

Now we want to transform the initial Hamiltonian ${\cal H}$ of
Eq.~(\ref{1}) into an effective Hamiltonian ${\cal H}_\lambda$
which has no matrix elements belonging to transitions larger than 
$\lambda$. This will be achieved by an unitary transformation  
\begin{eqnarray}
\label{6}
{\cal H}_\lambda &=& e^{X_\lambda}\; {\cal H}\;  e^{-X_\lambda} \; ,
 \end{eqnarray}
where the generator $X_\lambda$ of the transformation 
has to be anti-Hermitian, i.e. $X_\lambda^\dagger =- X_\lambda$. Note that the 
effective Hamiltonian ${\cal H}_\lambda$ has the same eigenspectrum as the 
original Hamiltonian ${\cal H}$. The generator $X_\lambda$ has to be 
chosen such that ${\cal H}_\lambda$ has no matrix
elements between states belonging to transitions larger than 
$\lambda$. Hence the condition 
\begin{eqnarray}
\label{7}
{\bf Q}_\lambda {\cal H}_\lambda &=& 0
 \end{eqnarray}
has to be fulfilled. Eq.~(\ref{7}) will be used below to 
specify $X_\lambda$. 

Instead of eliminating all  high-energy excitations in one step a sequence of 
stepwise transformations will be used in the following. 
Thus, in an infinitesimal
formulation, the projector-based renormalization approach yields 
renormalization equations for the parameters of the 
Hamiltonian as function of the cutoff $\lambda$. In that aspect 
the present approach resembles Wegner's flow equation 
method \cite{Wegner}     
and the similarity transformation introduced by  
G{\l}atzek and Wilson \cite{Glatzek}. 
To find the renormalization equations we proceed as follows. We start from the
renormalized Hamiltonian 
\begin{eqnarray}
\label{8}
{\cal H}_\lambda &=& {\cal H}_{0,\lambda} + {\cal H}_{1,\lambda} \,,
 \end{eqnarray}
where in  ${\cal H}_\lambda$ all excitations with energy 
differences larger than $\lambda$
have been eliminated. Next we further integrate out all excitations inside 
an energy shell between $\lambda$ and a smaller cutoff 
$(\lambda - \Delta \lambda)$ where $\Delta \lambda > 0$. The 
new Hamiltonian ${\cal H}_{(\lambda - \Delta \lambda)}$ 
is given by 
\begin{eqnarray}
\label{9}
{\cal H}_{(\lambda -\Delta \lambda)}
&=& e^{X_{\lambda, \Delta \lambda}} \; 
{\cal H}_{\lambda} \; e^{-X_{\lambda, \Delta \lambda}}\,,
 \end{eqnarray}
where $X_{\lambda, \Delta \lambda}$ is the generator for the 
transformation from $\lambda$ to $(\lambda - \Delta \lambda)$. 
Similar to (\ref{7}) it has to fulfill the condition 
\begin{eqnarray}
\label{10}
{\bf Q}_{(\lambda - \Delta \lambda)} \; 
{\cal H}_{(\lambda - \Delta \lambda)} &=& 0\,.
\end{eqnarray}
Note that there are two strategies to exploit Eq.~\eqref{10} in order 
to determine the
generator $X_{\lambda, \Delta \lambda}$ of the unitary transformation
\eqref{9}. The most straightforward route is to analyze Eqs.\eqref{9} and
\eqref{10} in perturbation theory as it was done in Refs. \onlinecite{Becker}
and \onlinecite{Hubsch1}. Here, we want to perform the renormalization step 
from $\lambda$ to $(\lambda - \Delta\lambda)$ in a non-perturbative way  
which recently has been applied to the periodic Anderson model 
by two of the authors\cite{Hubsch2}.  

Eqs.~(\ref{9}) and (\ref{10}) can be used to derive
difference equations for the $\lambda$ dependence of the 
parameters of the Hamiltonian.  They will be called renormalization 
equations. The solution depends on the initial values of the 
parameters of the Hamiltonian and fixes the final Hamiltonian 
${\cal H}_\lambda$ in the limit $ \lambda \rightarrow 0$. 
Note that the final Hamiltonian only consists of an
renormalized  unperturbed part 
${\cal H}_{0,(\lambda \rightarrow 0)}$. The interaction 
${\cal H}_{1,(\lambda \rightarrow 0)}$ completely vanishes since it was used 
up in the renormalization procedure.

\subsection{Application to  the Holstein model}

Let us start by formally writing down the effective 
Hamiltonian ${\cal H}_\lambda = e^{X_\lambda} {\cal H} e^{-X_\lambda}$
for the spinless Holstein model after all excitations 
with energy differences larger 
than $\lambda$ have been eliminated, 
\begin{eqnarray*}
{\cal H}_\lambda &=& {\cal H}_{0,\lambda}
+ {\cal H}_{1,\lambda}
\end{eqnarray*}
with 
 \begin{eqnarray}
\label{11}
{\cal H}_{0,\lambda } &=& 
\sum_{{\bf k}  } \varepsilon_{{\bf k}, \lambda} \;
c_{{\bf k} }^\dagger c_{{\bf k}}
+ \sum_{\bf q} \omega_{{\bf q},\lambda} \;b_{\bf q}^\dagger b_{\bf q}
+E_\lambda\,, \\
&& \nonumber \\
\label{12}
{\cal H}_{1,\lambda} &=& \sum_{{\bf k}, {\bf q}}
\frac{g_{{\bf k}, {\bf q}, \lambda}}{\sqrt N} \;
{\bf P}_\lambda\; 
\left(
  b_{\bf q}^\dagger c_{{\bf k}}^\dagger 
  c_{{\bf k}+{\bf q}} +
  b_{\bf q} c_{{\bf k}+ {\bf q}}^\dagger c_{{\bf k} } 
\right) \, ,
\end{eqnarray}
where Fourier transformed operators
\begin{eqnarray*}
  c_{{\bf k}}^\dagger &=& 
  \frac{1}{\sqrt{N}}\sum_{j} e^{i{\bf k}{\bf R}_{j}} c_{j}^\dagger,
  \qquad\qquad
  b_{{\bf q}}^\dagger \,=\, 
  \frac{1}{\sqrt{N}}\sum_{j} e^{i{\bf q}{\bf R}_{j}} b_{j}^\dagger,
\end{eqnarray*}
were introduced. 
Due to the renormalization processes the one-particle energies
$\varepsilon_{{\bf k}, \lambda}$ and the phonon frequencies
$\omega_{{\bf q}, \lambda}$ in (\ref{11}) now depend on the 
cutoff $\lambda$. Moreover, the phonon energies acquire a dispersion 
due to an effective interaction of lattice vibrations at different sites
via the coupling to electronic degrees of freedom.
Also the electron-phonon coupling constant $g_{{\bf k}, {\bf q}, \lambda}$
now depends on wave vectors ${\bf k}, {\bf q}$ and the cutoff $\lambda$. 
$E_\lambda$ 
is an additional energy shift. Finally, the projector ${\bf P}_\lambda$
in Eq.~(\ref{12}) guarantees that only those excitations
survive  in  ${\cal H}_{1,\lambda}$ which have energies (with
respect to ${\cal H}_{0,\lambda}$) smaller 
than $\lambda$: 
\begin{eqnarray}
\label{13}
\lefteqn{
  {\bf P}_\lambda\; 
  \left(
    b_{\bf q}^\dagger c_{{\bf k}}^\dagger 
    c_{{\bf k}+{\bf q}} +
    b_{\bf q} c_{{\bf k}+ {\bf q}}^\dagger c_{{\bf k} } 
  \right)
  \,=\,
} && \\ 
&=& 
\Theta\left(
  \lambda - 
  \left|
    \omega_{{\bf q},\lambda} + \varepsilon_{{\bf k},\lambda}
    -\varepsilon_{{\bf k}+{\bf q}, \lambda}
  \right| 
\right) \;
\left(
  b_{\bf q}^\dagger c_{{\bf k}}^\dagger 
  c_{{\bf k}+{\bf q}} +
  b_{\bf q} c_{{\bf k}+ {\bf q}}^\dagger c_{{\bf k} } 
\right)\,.
\nonumber
 \end{eqnarray}
The initial values of the 
original model (with cutoff $\lambda= \Lambda$) are 
\begin{eqnarray}
\label{14}
\varepsilon_{{\bf k}, (\lambda=\Lambda)} &=& \varepsilon_{\bf k}\,,
\hspace*{1cm} \omega_{{\bf q}, (\lambda =\Lambda)}= \omega_0\,,
\hspace*{1cm} g_{{\bf k},{\bf q}, (\lambda= \Lambda)}=
g\,, \hspace*{1cm} E_{(\lambda= \Lambda)}=0 \,.
\end{eqnarray}

In the next step we determine the renormalized 
Hamiltonian ${\cal H}_{(\lambda - \Delta \lambda)}
=e^{X_{\lambda, \Delta \lambda}} {\cal H}_\lambda 
e^{-X_{\lambda, \Delta \lambda}}$
by eliminating
all excitations within an additional  
small energy shell between $(\lambda -\Delta
\lambda)$ and $\lambda$.  For the explicit form of the 
generator $X_{\lambda, \Delta \lambda}$ of the unitary transformation     
we make the following ansatz 
\begin{eqnarray}
\label{15}
X_{\lambda, \Delta \lambda} &=& 
\frac{1}{\sqrt N} \sum_{{\bf k}, {\bf q}} 
B_{{\bf k}, {\bf q}, \lambda} \; 
\Theta_{{\bf k}, \bf q}(\lambda, \Delta \lambda) \;
\left(
  b_{\bf q}^\dagger c_{{\bf k}}^\dagger 
  c_{{\bf k}+{\bf q}} -
  b_{\bf q} c_{{\bf k}+ {\bf q}}^\dagger c_{{\bf k} } 
\right)\,,
\end{eqnarray}
where $\Theta_{{\bf k}, \bf q}(\lambda, \Delta \lambda)$
is the product of two $\Theta$-functions
\begin{eqnarray}
\label{16}
\Theta_{{\bf k}, \bf q}(\lambda, \Delta \lambda) &=&
\Theta(\lambda -|\omega_{{\bf q},\lambda} + \varepsilon_{{\bf k},\lambda}
-\varepsilon_{{\bf k}+{\bf q}, \lambda}| ) \\ 
&& \times 
\Theta\left[
  |\omega_{{\bf q},(\lambda-\Delta \lambda)} + 
  \varepsilon_{{\bf k},(\lambda- \Delta \lambda)}
  -\varepsilon_{{\bf k}+{\bf q}, (\lambda- \Delta \lambda)}| -(\lambda 
  -\Delta \lambda) 
\right]\,. \nonumber
\end{eqnarray}
The operator form of $X_{\lambda, \Delta \lambda}$ is suggested by its 
first order expression which is easily obtained by expanding (\ref{9})
in powers of ${\cal H}_1$ and using (\ref{10}), (cf. Ref.~\onlinecite{Becker}).
The yet unknown prefactors $B_{{\bf k}, {\bf q}, \lambda}$
will be specified later and depend on $\lambda$. It will turn out that 
$B_{{\bf k}, {\bf q}, \lambda}$
contains contributions in all powers of the electron-phonon 
coupling $g$. The two $\Theta$-functions in (\ref{16}) confine the allowed
excitations to the energy shell $\Delta \lambda$.

\bigskip
The coefficients
$B_{{\bf k},{\bf q},\lambda}$ will be fixed by the condition (\ref{10}).
First, we have to carry out the unitary transformation 
(\ref{9}) explicitly:
\begin{eqnarray}
\label{17}
{\cal H}_{(\lambda -\Delta \lambda)} &=&
\sum_{{\bf k}} \varepsilon_{{\bf k},\lambda}\;
e^{X_{\lambda,\Delta \lambda}} c_{{\bf k}}^\dagger c_{{\bf k}}
e^{-X_{\lambda,\Delta \lambda}} +
\sum_{{\bf q}} \omega_{{\bf q},\lambda} \;
e^{X_{\lambda,\Delta \lambda}} b_{{\bf q}}^\dagger b_{{\bf q}}
e^{-X_{\lambda,\Delta \lambda}} + E_\lambda + \nonumber \\
&&+
 \sum_{{\bf k}, {\bf q}}
\frac{g_{{\bf k}, {\bf q}, \lambda}}{\sqrt N} \;
e^{X_{\lambda,\Delta \lambda}}
{\bf P}_\lambda\; 
\left(
  b_{\bf q}^\dagger c_{{\bf k}}^\dagger 
  c_{{\bf k}+{\bf q}} +
  b_{\bf q} c_{{\bf k}+ {\bf q}}^\dagger c_{{\bf k} } 
\right)\, 
e^{-X_{\lambda,\Delta \lambda}}\,.  
\end{eqnarray}
The transformations for the various operators in (\ref{17}) has to be done
separately. For instance, the transformation for $c_{{\bf k}}^\dagger
c_{{\bf k}}$ reads 
\begin{eqnarray}
\label{18}
\lefteqn{
  e^{X_{\lambda,\Delta \lambda}} \, 
  c_{{\bf k}}^\dagger c_{{\bf k}} \,
  e^{-X_{\lambda,\Delta \lambda}} - 
  c_{{\bf k}}^\dagger c_{{\bf k}} \,=\,
} && \\[1ex]
&=& \sum_{\bf q} 
\left\{
  \left[ 
    \frac{\Theta_{{\bf k}-{\bf q},{\bf q}}(\lambda,\Delta\lambda)}
    {\hat{n}_{{\bf q},\lambda}}
    \left(
      \cos (
        B_{{\bf k}-{\bf q},{\bf q},\lambda} \sqrt{2\hat{n}_{{\bf q},\lambda}}
      ) - 1
    \right)
    \left\{ 
      ( 1 - n_{{\bf k}- {\bf q},\lambda}^c +n_{{\bf q}, \lambda}^b) \;
      c_{{\bf k}}^\dagger  c_{{\bf k}}
    \right.
  \right.
\right. \nonumber \\ 
&& \qquad \qquad
\left.
  \left.
    \left.
      - (n_{{\bf k},\lambda}^c +n_{{\bf q}, \lambda}^b) \;
      c_{{\bf k}-{\bf q}}^\dagger  c_{{\bf k}-{\bf q}}
    \right\}
  \right.
\right.
\nonumber \\
&& \qquad
\left.
  \left.
    +
    \frac{
      \Theta_{{\bf k}-{\bf q}, {\bf q}}(\lambda,\Delta\lambda) 
      (n_{{\bf k},\lambda}^c -n_{{\bf k}-
      {\bf q},\lambda}^c)
    }{\hat{n}_{{\bf q}, \lambda}} 
    \left\{
      \cos(B_{{\bf k}-{\bf q},{\bf q},\lambda}
      \sqrt{2\hat{n}_{{\bf q}, \lambda}}) - 1 
    \right\} \; 
    b_{\bf q}^\dagger b_{\bf q}
  \right.
\right. \nonumber \\
&& \qquad
\left.
  \left.
    +
    \frac{\Theta_{{\bf k}-{\bf q}, {\bf q}}(\lambda,\Delta\lambda) }  
    {\sqrt{2 \hat{n}_{{\bf q},\lambda}}} 
    \sin(B_{{\bf k}-{\bf q},{\bf q},\lambda}\sqrt{2\hat{n}_{{\bf q},\lambda}}) 
    \; 
    \left(
      b_{\bf q}^\dagger c_{{\bf k}-{\bf q}}^\dagger c_{{\bf k}} 
      + {\rm h.c.}
    \right)
  \right] \; 
  - \left[ {\bf k} \rightarrow ({\bf k} + {\bf q}) \right] 
\right\}\,,\nonumber
\end{eqnarray} 
where we have defined 
\begin{eqnarray}
\label{19}
n_{{\bf k},\lambda}^c &=& \big< c_{{\bf k}}^\dagger 
c_{{\bf k}} \big>_{0,\lambda}
= \frac{1}{e^{\beta \varepsilon_{{\bf k},\lambda}}+1} \ ,
\hspace*{2cm} 
n_{{\bf q},\lambda}^b = \big< b_{\bf q}^\dagger b_{\bf q} \big>_{0,\lambda}
= \frac{1} {e^{\beta \omega_{{\bf q},\lambda}} -1}\,,
\end{eqnarray}
and 
\begin{eqnarray}
\label{20}
\hat{n}_{{\bf q},\lambda} &=& 1 + 2n_{{\bf q},\lambda}^b 
\end{eqnarray}
(for details see Appendix \ref{A}). 
In deriving (\ref{18}) an additional
factorization approximation has been used in order to keep only operators 
of the structure of those of Eqs.~(\ref{11}) and (\ref{12}). 
In principle, the expectation values are {\bf best} 
defined with the equilibrium 
distribution of ${\cal H}_\lambda$ since the renormalization 
step was done from  ${\cal H}_\lambda$ to 
${\cal H}_{(\lambda - \Delta \lambda)}$. However, for simplicity
we shall evaluate these quantities with the unperturbed Hamiltonian 
${\cal H}_{0,\lambda}$. 
Similar expressions to (\ref{18}) are  also found for the
transformations of the remaining operators of (\ref{17}). 

One should note that 
influence of the above factorization 
approximation together with the choice of the 
equilibrium distribution is not well controlled 
as long as the influence of the additional fluctuation terms 
are not considered. However, we believe that 
the renormalization of ${\cal H}_\lambda$ as it is used here  
leads to a proper description of the influence 
of the electron-phonon interaction in the Holstein model. The 
fluctuation terms which appear as additional renormalization 
contributions in the approach could give rise to additional terms 
both in ${\cal H}_{0,\lambda}$ and ${\cal H}_{1,\lambda}$. 
The renormalization of the new coupling parameters  
would have to be investigated in the present renormalization scheme
as well.

In the next step we determine the parameters 
$B_{{\bf k},{\bf q},\lambda}$. For that purpose  we use the condition
(\ref{10}). By  inserting ${\cal H}_{(\lambda -
\Delta \lambda)}$ from (\ref{17}) into (\ref{10}) and by use of (\ref{18}) 
and the remaining transformations, we find 
\begin{eqnarray}
\label{21}
\Theta_{{\bf k},{\bf q}}
(\lambda, \Delta \lambda)  \;
B_{{\bf k}, {\bf q}, \lambda} &=& 
\Theta_{{\bf k},{\bf q}}(\lambda, \Delta \lambda) \;
\frac{1}{\sqrt{2\hat{n}_{{\bf q},\lambda}}}
\arctan\left(\sqrt{\frac{2\hat{n}_{{\bf q},\lambda}}{N}}
\frac{g_{{\bf k}, {\bf q},\lambda}}{\omega_{{\bf q},\lambda} 
+ \epsilon_{{\bf k},\lambda} 
- \epsilon_{{\bf k}+ {\bf q} ,\lambda} } \right)\,. 
\end{eqnarray}
Let us point out that $B_{{\bf k},{\bf q},\lambda}$ is determined by (\ref{21})
only for the case that the excitation energies 
$\omega_{{\bf q},\lambda} + \varepsilon_{{\bf k}, \lambda} - 
\varepsilon_{{\bf k}+{\bf q},\lambda}$ fit into the energy shell
given by $\Theta_{{\bf k},{\bf q}}(\lambda, \Delta \lambda)$. 
For all other excitations  $B_{{\bf k},{\bf q},\lambda}$ can be set equal 
to zero. Therefore  we shall use the following
expression for $B_{{\bf k},{\bf q},\lambda}$:
\[
\begin{array}{c}
B_{{\bf k}, {\bf q}, \lambda} = \left\{ 
\begin{array}{cc}
\displaystyle 
\frac{1}{\sqrt{\displaystyle 2\hat{n}_{{\bf q},\lambda}}}
\arctan\left(\sqrt{\frac{\displaystyle 2\hat{n}_{{\bf q},\lambda}}{N}}
\frac{\displaystyle g_{{\bf k}, {\bf q},\lambda}}{\omega_{{\bf q},\lambda} 
+ \epsilon_{{\bf k},\lambda} 
- \epsilon_{{\bf k}+ {\bf q} ,\lambda} } \right) 
& \hspace*{1cm} \mbox{for\ } 
\Theta_{{\bf k},{\bf q}}(\lambda, \Delta \lambda)  =1  \\ 
0   
& \hspace*{1cm} \mbox{for\ }  
\Theta_{{\bf k}, {\bf q}} (\lambda, \Delta \lambda) =0\,.
\end{array} \right .
\end{array}
\]

\subsection{Renormalization equations for the Holstein model}

Next, we derive the renormalization equations for the parameters of
the Hamiltonian. For that purpose we compare the renormalization
ansatz  for 
${\cal H}_{(\lambda - \Delta \lambda)}$,
\begin{eqnarray}
\label{22}
{\cal H}_{(\lambda-\Delta \lambda)} &=&
\sum_{{\bf k},  } \varepsilon_{{\bf k}, (\lambda-\Delta \lambda)} \;
c_{{\bf k} }^\dagger c_{{\bf k}}
+ \sum_{\bf q} \omega_{{\bf q},(\lambda-\Delta\lambda)} 
\;b_{\bf q}^\dagger b_{\bf q}
+E_{(\lambda-\Delta \lambda)} \\
&& \nonumber \\
 &+& \sum_{{\bf k}, {\bf q}}
\frac{g_{{\bf k}, {\bf q}, (\lambda-\Delta\lambda)}}{\sqrt N} \;
{\bf P}_{(\lambda -\Delta \lambda)}
\; 
\left(
  b_{\bf q}^\dagger c_{{\bf k}}^\dagger 
  c_{{\bf k}+{\bf q}} +
  b_{\bf q} c_{{\bf k}+ {\bf q}}^\dagger c_{{\bf k} } 
\right) 
\nonumber
 \end{eqnarray}
[see Eqs.~(\ref{11}) and (\ref{12})], with the expression that is obtained
from (\ref{17}) after (\ref{18}) and the corresponding 
transformations have been inserted. 
Comparing in both equations the coefficients of
the operators $c_{{\bf k}}^\dagger c_{{\bf k} }$,
$b_{\bf q}^\dagger b_{\bf q}$, and 
$(b_{\bf q}^\dagger c_{{\bf k}}^\dagger c_{{\bf k}+{\bf q}}  
+h.c.)$, we find the following relations between the parameters
at cutoff $\lambda$ and $(\lambda -\Delta \lambda)$:
\begin{eqnarray}
\label{23}
\lefteqn{
  \varepsilon_{{\bf k},(\lambda - \Delta\lambda)} - 
  \varepsilon_{{\bf k},\lambda}
  \,=\,
} &&\\
&=& 
\sum_{\bf q} \frac{n_{{\bf k}+{\bf q},\lambda}^{c} + n_{{\bf q},\lambda}^{b}}
{\hat{n}_{{\bf q},\lambda}} \left\{\cos(B_{{\bf k}, {\bf q},\lambda}
\sqrt{2\hat{n}_{{\bf q},\lambda}}) - 1 \right\} 
\left(\omega_{{\bf q},\lambda} + \varepsilon_{{\bf k},\lambda} 
- \varepsilon_{{\bf k} +{\bf q},\lambda} \right) 
\Theta_{{\bf k},{\bf q}}(\lambda,\Delta\lambda) \nonumber\\ 
&&
+ \sum_{\bf q}
\sqrt{\frac{2}{N \hat{n}_{{\bf q},\lambda}}} 
\left(n_{{\bf k} +{\bf q},\lambda}^{c} + n_{{\bf q},\lambda}^{b} \right)
\sin(B_{{\bf k},{\bf q},\lambda} 
\sqrt{2\hat{n}_{{\bf q},\lambda}}) g_{{\bf k},{\bf q},\lambda} 
\Theta_{{\bf k},{\bf q}}(\lambda,\Delta\lambda) \nonumber\\ 
&&
-  \sum_{\bf q} \frac{1 - n_{{\bf k}-{\bf q},\lambda}^{c} 
+ n_{{\bf q},\lambda}^{b}}{\hat{n}_{{\bf q},\lambda}} 
\left\{\cos(B_{{\bf k}-{\bf q},{\bf q},\lambda}
\sqrt{2\hat{n}_{{\bf q},\lambda}}) 
- 1 \right\} \left(\omega_{{\bf q},\lambda} + 
\varepsilon_{{\bf k}-{\bf q},\lambda} 
- \varepsilon_{{\bf k},\lambda} \right) 
\Theta_{{\bf k}-{\bf q},{\bf q}}(\lambda,\Delta\lambda) \nonumber\\ 
&&
- \sum_{\bf q}
\sqrt{\frac{2}{N \hat{n}_{{\bf q},\lambda}}} 
\left(1 - n_{{\bf k}-{\bf q},\lambda}^{c} + n_{{\bf q},\lambda}^{b} \right)
\sin(B_{{\bf k}-{\bf q},{\bf q},\lambda} 
\sqrt{2\hat{n}_{{\bf q},\lambda}}) g_{{\bf k}-
{\bf q},{\bf q},\lambda} 
\Theta_{{\bf k}-{\bf q},{\bf q}}(\lambda,\Delta\lambda)\,, \nonumber\\[4ex]
\label{24}
\lefteqn{
  \omega_{{\bf q},(\lambda - \Delta\lambda)} - \omega_{{\bf q},\lambda} 
  \,=\,
}
&&\\
&=&
- \sum_{\bf k} \frac{n_{{\bf k}+{\bf q},\lambda}^{c} - 
n_{{\bf k},\lambda}^{c}}{\hat{n}_{{\bf q},\lambda}} 
\left\{\cos(B_{{\bf k},{\bf q},\lambda}
\sqrt{2\hat{n}_{{\bf q},\lambda}}) - 1 \right\} 
\left(\omega_{{\bf q},\lambda} + \varepsilon_{{\bf k},\lambda} 
- \varepsilon_{{\bf k}+{\bf q},\lambda} \right) 
\Theta_{{\bf k},{\bf q}}(\lambda,\Delta\lambda) \nonumber\\ 
&&
-\sum_{\bf k}
\sqrt{\frac{2}{N \hat{n}_{{\bf q},\lambda}}} 
\left(n_{{\bf k}+{\bf q},\lambda}^{c} - n_{{\bf k},\lambda}^{c} \right)
\sin(B_{{\bf k},{\bf q},\lambda} \sqrt{2\hat{n}_{{\bf q},\lambda}}) \; 
g_{{\bf k},{\bf q},\lambda} 
\Theta_{{\bf k},{\bf q}}(\lambda,\Delta\lambda)\,,\nonumber\\[4ex]
\label{25}
\lefteqn{
  g_{{\bf k},{\bf q},(\lambda - \Delta\lambda)} -
  g_{{\bf k},{\bf q},\lambda}
  \,=\,
}\\
&=&
-\sqrt{\frac{N}{2\hat{n}_{{\bf q},\lambda}}} 
\left(\omega_{{\bf q},\lambda} + \varepsilon_{{\bf k},\lambda} 
- \varepsilon_{{\bf k}+{\bf q},\lambda} \right) 
\sin(B_{{\bf k},{\bf q},\lambda} \sqrt{2\hat{n}_{{\bf q},\lambda}}) \; 
\Theta_{{\bf k},{\bf q}}(\lambda,\Delta\lambda)\nonumber \\
&&
+ g_{{\bf k},{\bf q},\lambda} 
\left\{\cos(B_{{\bf k},{\bf q},\lambda}
\sqrt{2\hat{n}_{{\bf q},\lambda}}) - 1 \right\}  
\Theta_{{\bf k},{\bf q}}(\lambda,\Delta\lambda)\,.\nonumber 
\end{eqnarray}
A corresponding expression can also be found for 
the renormalization 
of the energy shift $E_{(\lambda -\Delta \lambda)}$. 
The above Eqs.~(\ref{23}) to 
(\ref{25}) describe the renormalization of the parameters 
if the cutoff is reduced from $\lambda$ to $(\lambda -\Delta \lambda)$. 

Note that the expression (\ref{21}) for $B_{{\bf k},{\bf q},\lambda}$
and thus the renormalization equations  are
non-perturbative in $g_{{\bf k},{\bf q}, \lambda}$
and are not restricted to some low 
order. However, a prefactor $1/\sqrt{N}$ enters 
the expression (\ref{21}) for $B_{{\bf k},{\bf q},\lambda}$ due to the 
factorization approximation discussed in  
Appendix A. Thus, in the thermodynamic 
limit $N\rightarrow \infty$, $B_{{\bf k},{\bf q},\lambda}$ 
becomes linear in $g$ and the renormalization relations (\ref{23}) 
to (\ref{25}) become quadratic in the electron-phonon coupling. 
In the numerical evaluation, however,  
when a fixed number $N$ is taken, the excitation 
energy $(\omega_{{\bf q},\lambda} + \varepsilon_{{\bf k},\lambda}
-\varepsilon_{{\bf k}+{\bf q}, \lambda})$ will become very small 
for reducing the cutoff $\lambda \rightarrow 0$ and thus  the expansion 
of $B_{{\bf k},{\bf q},\lambda}$ to linear order in $g$ may break down. In that
case  the full equations (\ref{21}) and (\ref{23}) to (\ref{25})
have to be taken.    

The overall renormalization starts from the cutoff $\lambda = \Lambda$ 
of the original model and proceeds down to  zero cutoff $\lambda=0$. At 
$\lambda =0$ the 
completely renormalized Hamiltonian $\tilde{\cal H}:=
{\cal H}_{(\lambda \rightarrow 0)}$ becomes an effectively free model 
and reads
\begin{eqnarray}
\label{26}
\tilde{\cal H} &=&
\sum_{{\bf k},  } \tilde{\varepsilon}_{{\bf k}} \;
c_{{\bf k} }^\dagger c_{{\bf k}}
+ \sum_{\bf q} \tilde{\omega}_{{\bf q}}
\;b_{\bf q}^\dagger b_{\bf q}
+\tilde{E}\,,  
\end{eqnarray}
where we have defined $\tilde{\varepsilon}_{\bf k} =
\varepsilon_{{\bf k}, (\lambda \rightarrow 0)}, 
\tilde{\omega}_{\bf q} = \omega_{{\bf q},(\lambda \rightarrow 0)}$,
and $\tilde{E} = E_{(\lambda \rightarrow 0)}$. For $\lambda 
\rightarrow 0$ the electron-phonon coupling 
${\cal H}_{1,\lambda}$ has vanished due to 
the $\Theta$-function in \eqref{13}. All excitations of the 
electron-phonon interaction were used to renormalize the 
parameters of the Hamiltonian.

\bigskip
The results from the numerical evaluation of the 
renormalization equations will be given in Sect.~IV. 
In particular we then compare the spectral functions 
calculated within the PRM approach and by exact diagonalization. 
The electronic one-particle  spectral functions  
$A_{\bf k}^+(\omega)$ and  $A_{\bf k}^-(\omega)$
are defined by   
\begin{eqnarray}
\label{27}
A_{\bf k}^+(\omega) &=& \frac{1}{2\pi} \int_{-\infty}^{\infty}
\big< c_{{\bf k} } (t) \;c_{{\bf k}}^\dagger
\big> \; e^{i\omega t} dt\,, \hspace*{1cm}
A_{\bf k}^-(\omega) = \frac{1}{2\pi} \int_{-\infty}^{\infty}
\big< c_{{\bf k} }^\dagger (t) \;c_{{\bf k}}
\big> \; e^{i\omega t} dt\,.    
\end{eqnarray}
The function $A_{\bf k}^+(\omega)$ describes the creation of an electron 
${\bf k}$ at time zero and the annihilation at time $t$ whereas in
$A_{\bf k}^-(\omega)$ first an electron is annihilated and at time $t$ 
the electron is created. As is well-known these quantities can be measured 
by inverse photoemission (IPE) and by photoemission (PE). 
To evaluate $A_{\bf k}^+(\omega)$ and $A_{\bf k}^-(\omega)$
 within the PRM approach it is necessary to
unitary transform not only the Hamiltonian but also the 
operators $c_{{\bf k} }$ and  $c_{{\bf k} }^\dagger$.
This follows from the fact that the trace of any operator quantity
is invariant under a  unitary transformation. Thus we have to evaluate 
\begin{eqnarray}
\label{28}
A_{\bf k}^+(\omega) &=& \frac{1}{2 \pi}
\int_{-\infty}^{\infty} \big< (c_{{\bf k}})_\lambda(t)
(c_{{\bf k} }^\dagger)_\lambda \big>_\lambda \;
e^{i\omega t} dt\,,
\end{eqnarray}
where the expectation value and the time dependence are defined 
with respect to the $\lambda$-dependent Hamiltonian 
${\cal H}_\lambda$. Moreover,  
$  (c_{{\bf k} }^\dagger)_\lambda = e^{X_{\lambda}}
c_{{\bf k} }^\dagger  e^{-X_{\lambda}}$. A similar
expression to Eq.~\eqref{28} is also valid for $A_{\bf k}^-(\omega)$.

In analogy to the evaluation procedure for the 
renormalization equations of ${\cal H}_\lambda$ 
we make the following ansatz for the $\lambda$-dependence of the  operators   
$(c_{{\bf k} }^\dagger)_\lambda$
and $(c_{{\bf k} })_\lambda$,
\begin{eqnarray}
\label{29}
 (c_{{\bf k} }^\dagger)_\lambda &=&
\alpha_{{\bf k},\lambda} \; c_{{\bf k}}^\dagger
+ \sum_{\bf q} \big(\beta_{{\bf k},{\bf q},\lambda} \;
c_{{\bf k}+{\bf q}}^\dagger b_{\bf q} + 
\gamma_{{\bf k},{\bf q},\lambda} \; c_{{\bf k}-{\bf q}}^\dagger
b_{\bf q}^\dagger \big) \ , 
\end{eqnarray}
$ (c_{{\bf k}})_\lambda=
[(c_{{\bf k}})_\lambda^\dagger]^\dagger $,
with $\lambda$-dependent parameters $\alpha_{{\bf k}\lambda}$,
$\beta_{{\bf k},{\bf q},\lambda}$, and $\gamma_{{\bf k},{\bf q},\lambda}$.
The parameter values for the original model are 
\begin{eqnarray}
\label{30}
 \alpha_{{\bf k},(\lambda= \Lambda)} &=& 1\,, \hspace*{1cm}
\beta_{{\bf k},{\bf q},(\lambda=\Lambda)} =0\,, \hspace*{1cm}
\gamma_{{\bf k},{\bf q},(\lambda=\Lambda)} = 0\,.
\end{eqnarray}
Due to \eqref{29}  the following sum rule must hold
\begin{eqnarray}
\label{31}
 1 \,=\, |\alpha_{{\bf k},\lambda}|^2 &+& 
\sum_{\bf q} 
\left\{
  |\beta_{{\bf k},{\bf q}, \lambda}|^2 
  \left( 
    n_{{\bf k}+{\bf q},\lambda}^c
    + 
    n_{{\bf q},\lambda}^b
  \right) +
\right.
\nonumber \\
&& \quad
\left.
  +|\gamma_{{\bf k},{\bf q},\lambda}|^2 
  \left( 
    1 + n_{{\bf q},\lambda}^b - 
    n_{{\bf k}-{\bf q},\lambda}^c
  \right)
\right\}\,,
\end{eqnarray}
which follows from the commutator relations. Note that a 
factorization approximation was used on the right hand side
of \eqref{31}. 
In analogy to the former approach
for $\varepsilon_{{\bf k},\lambda}, \omega_{{\bf q}, \lambda}$, 
and $g_{{\bf k},{\bf q}, \lambda}$, we find the following 
renormalization equations for the parameters  $\alpha_{{\bf k}\lambda}$,
$\beta_{{\bf k},{\bf q},\lambda}$ and $\gamma_{{\bf k},{\bf q},\lambda}$:
\begin{eqnarray}
\label{32}
\lefteqn{
  \alpha_{{\bf k}, \lambda -\Delta \lambda} - \alpha_{{\bf k},\lambda}
  \,=\,
}&&\\
&=&
\sum_{\bf q}
\left\{ 
  \cos\left(
    \sqrt{ n_{{\bf k}+{\bf q},\lambda}^c + n_{{\bf q},\lambda}^b}\; 
    B_{{\bf k},{\bf q},\lambda}
  \right) - 1 
\right\} 
\; \alpha_{{\bf k}, \lambda} 
\Theta_{{\bf k},{\bf q}}(\lambda,\Delta\lambda) \nonumber \\
&&
+ \sum_{\bf q} 
\left\{ 
  \cos\left(
    \sqrt{ 1- n_{{\bf k}-{\bf q},\lambda}^c + n_{{\bf q},\lambda}^b}\; 
    B_{{\bf k}-{\bf q},{\bf q},\lambda}
  \right) - 1 
\right\} 
\; \alpha_{{\bf k}, \lambda} 
\Theta_{{\bf k}-{\bf q},{\bf q}}(\lambda,\Delta\lambda) \nonumber \\
&&
+ \sum_{\bf q} \sqrt{n_{{\bf k}+{\bf q},\lambda}^c + n_{{\bf q},\lambda}^b} \;
\sin\left(
  \sqrt{ n_{{\bf k}+{\bf q},\lambda}^c + n_{{\bf q},\lambda}^b}\; 
  B_{{\bf k},{\bf q},\lambda}
\right)
\; \beta_{{\bf k},{\bf q},  \lambda} 
\Theta_{{\bf k},{\bf q}}(\lambda,\Delta\lambda) \nonumber \\
&&
- \sum_{\bf q} 
\sqrt{1- n_{{\bf k}- {\bf q},\lambda}^c + n_{{\bf q},\lambda}^b} \;
\sin\left(
  \sqrt{ 1- n_{{\bf k}-{\bf q},\lambda}^c + n_{{\bf q},\lambda}^b}\; 
  B_{{\bf k}-{\bf q},{\bf q},\lambda}
\right)
\; \gamma_{{\bf k},{\bf q},  \lambda} 
\Theta_{{\bf k}-{\bf q},{\bf q}}(\lambda,\Delta\lambda)\,, \nonumber \\[4ex]
\label{33}
\lefteqn{
  \beta_{{\bf k}, {\bf q}, \lambda -\Delta \lambda} -
  \beta_{{\bf k}, {\bf q}, \lambda}
  \,=\,
} &&\\
&=& 
- \frac{1}{ \sqrt{n_{{\bf k}+{\bf q},\lambda}^c + n_{{\bf q},\lambda}^b}}
\sin \left(
  \sqrt{n_{{\bf k}+{\bf q},\lambda}^c 
  + n_{{\bf q},\lambda}^b}\; B_{{\bf k},{\bf q},\lambda}
\right) \; 
\alpha_{{\bf k}, \lambda} \Theta_{{\bf k}, {\bf q}}(\lambda,\Delta\lambda)
\nonumber \\
&&
+ \left\{
  \cos \left( 
    \sqrt{n_{{\bf k}+{\bf q},\lambda}^c 
    + n_{{\bf q},\lambda}^b}\; B_{{\bf k},{\bf q},\lambda}
  \right) - 1
\right\} \; 
\beta_{{\bf k},{\bf q}, \lambda} 
\Theta_{{\bf k}, {\bf q}}(\lambda,\Delta\lambda)\,,\nonumber
\end{eqnarray}
and 
\begin{eqnarray} 
\label{34}
\lefteqn{
  \gamma_{{\bf k}, {\bf q}, \lambda -\Delta \lambda} - 
  \gamma_{{\bf k}, {\bf q}, \lambda}
  \,=\,
}&&\\
&=& 
\frac{1}{ \sqrt{1- n_{{\bf k}-{\bf q},\lambda}^c + n_{{\bf q},\lambda}^b}}
\sin \left( 
  \sqrt{1- n_{{\bf k}-{\bf q},\lambda}^c + n_{{\bf q},\lambda}^b}\; 
  B_{{\bf k}-{\bf q},{\bf q},\lambda}
\right) \; 
\alpha_{{\bf k}, \lambda} 
\Theta_{{\bf k}-{\bf q}, {\bf q}}(\lambda,\Delta\lambda)
\nonumber \\
&&
+ \left\{
  \cos \left(
    \sqrt{1- n_{{\bf k}-{\bf q},\lambda}^c + n_{{\bf q},\lambda}^b}\; 
    B_{{\bf k}-{\bf q},{\bf q},\lambda}
  \right) - 1 
\right\} \; 
\gamma_{{\bf k},{\bf q}, \lambda} 
\Theta_{{\bf k}-{\bf q}, {\bf q}}(\lambda,\Delta\lambda)\,.\nonumber
\end{eqnarray}
The renormalization equations (\ref{31}) and (\ref{33}) 
can again be integrated numerically by reducing the 
cutoff $\lambda$ stepwise down to $\lambda \rightarrow 0$. 
Together with the Eqs.~(\ref{23}) to  (\ref{25}) the 
spectral functions (\ref{27}) can be determined. 
\begin{eqnarray} 
\label{35}
A_{{\bf k}}^+(\omega) &=& 
\tilde{\alpha}_{\bf k}^2 \; \delta(\omega - \tilde{\varepsilon}_{\bf k} )
(1- \tilde{n}_{\bf k}^c) + 
\sum_{\bf q} \tilde{\beta}_{{\bf k}, {\bf q}}^2 \;
\delta(\omega +\tilde{\omega}_{\bf q} - \tilde{\varepsilon}_{{\bf k}+{\bf q}})
\; \tilde{n}_{{\bf q}}^b (1 - \tilde{n}_{{\bf k} +{\bf q}}^c) \nonumber \\
&& + \sum_{\bf q} \tilde{\gamma}_{{\bf k}, {\bf q}}^2\;
\delta(\omega - \tilde{\omega}_{\bf q} - 
\tilde{\varepsilon}_{{\bf k}-{\bf q}})\: (1-\tilde{n}_{{\bf k} -{\bf q}}^c)
(1+ \tilde{n}_{\bf q}^b)\, , \\
&& \nonumber \\
\label{36}
A_{{\bf k}}^-(\omega) &=& 
\tilde{\alpha}_{\bf k}^2 \; \delta(\omega - \tilde{\varepsilon}_{\bf k} )
\tilde{n}_{{\bf k}}^c + 
\sum_{\bf q} \tilde{\beta}_{{\bf k}, {\bf q}}^2 \;
\delta(\tilde{\omega}_{\bf q} - \tilde{\varepsilon}_{{\bf k}+{\bf q}}
+\omega)\; \tilde{n}_{{\bf k}+{\bf q}}^c(1 + \tilde{n}_{\bf q}^b) \nonumber \\
&& + \sum_{\bf q} \tilde{\gamma}_{{\bf k}, {\bf q}}^2\;
\delta(\omega - \tilde{\omega}_{\bf q} - 
\tilde{\varepsilon}_{{\bf k}-{\bf q}})\: 
\tilde{n}_{{\bf k} -{\bf q}}^c \tilde{n}_{\bf q}^b \,.
\end{eqnarray}
As before, the quantities with tilde denote the parameters at 
cutoff $\lambda \rightarrow 0$
\begin{eqnarray}
\label{37}
\tilde{\alpha}_{\bf k} &=& \alpha_{{\bf k}, (\lambda \rightarrow 0)}\,
\hspace*{1cm}
\tilde{\beta}_{{\bf k}, {\bf q}} = 
\beta_{{\bf k}, {\bf q}, ( \lambda  \rightarrow 0 )} \,
\hspace*{1cm} 
\tilde{\gamma}_{{\bf k}, {\bf q}} =  
\gamma_{{\bf k}, {\bf q},(\lambda \rightarrow 0)}\,.
\end{eqnarray}
In deriving (\ref{35}) and 
(\ref{36}) it was exploited
that the completely renormalized Hamiltonian 
$\tilde{\cal H}$ is diagonal [cf. Eq.~(\ref{26})]. Note that the first 
parts in  $A_{\bf k}^+(\omega)$ and $A_{\bf k}^-(\omega)$ describe 
the coherent one-electron excitations 
which correspond to those of  a free electron  gas. 
The two remaining contributions  
are incoherent excitations due to the coupling to phonons 
and are given by a ${\bf q}$-sum  over excitations 
$\tilde{\varepsilon}_{{\bf k} \pm {\bf q}} \mp \tilde{\omega}_{\bf q}$.  
Finally we give  a sum rule which follows from the frequency integral 
over  the sum of the two spectral functions 
\begin{eqnarray}
\label{38}
1 &=& \int_{-\infty}^{\infty} d\omega
(A_{\bf k}^+(\omega) + A_{\bf k}^-(\omega) ) = \nonumber \\
&=& \tilde{\alpha}_{\bf k}^2 +
\sum_{\bf q} \tilde{\beta}_{{\bf k},{\bf q}}^2
(\tilde{n}_{{\bf k}+{\bf q}}^c + \tilde{n}_{\bf q}^b) +
\sum_{\bf q} \tilde{\gamma}_{{\bf k},{\bf q}}^2
(1 + \tilde{n}_{\bf q}^b - \tilde{n}_{{\bf k} -{\bf q}}^c )\,. 
\end{eqnarray}
Note that Eq.~(\ref{38}) is equivalent to the former 
relation (\ref{31}). The explicit evaluation of the 
spectral functions will  be performed in Sect.~IV below.

\section{Exact diagonalization of electron-phonon models}

In principle exact diagonalization techniques
allow the analysis of ground-state and spectral properties 
of microscopic models free of any approximations.
However, the vast Hilbert space dimensions 
restrict ED studies to rather small system sizes, even if state 
of the art supercomputers are used.
Usually the dimension of the matrix involved in the diagonalization 
is reduced by exploiting lattice or spin symmetries~\cite{BWF98}. 
Unfortunately, for the Holstein model the Hilbert space 
associated to the phonons 
is infinite even for finite systems. Thus  
a well controlled truncation procedure has to be developed~\cite{II90,BWF98}. 
In our approach the maximum number of phonons per state 
(not per lattice site!) is fixed ($M$). Then the 
dimension of the phononic subspace is $D_{ph}=\frac{(M+N)!}{M!N!}$. 

For the Holstein model computational requirements can be further reduced.  
It is possible to separate the symmetric 
 phonon mode, $B_{0}=\frac{1}{\sqrt{N}}\sum_{i} b_i$, and to
calculate its contribution to $H$ analytically.
For the sake of simplicity, we restrict ourselves to the 1D case in what follows.
Using the momentum space representation of the phonon operators 
and introducing the polaron binding energy $\varepsilon_p=g^2/\omega_0$
the original Holstein Hamiltonian reads
\begin{equation}
{\cal H} = -t \sum_{i,j} (c^{\dagger}_{i}c_{j}
 +{\rm h.c.}) - \sqrt{\varepsilon_p\omega_0} \;\sum_{j}(
{B^{\dagger}_{-Q_j}}+ {B^{}_{Q_j}}) {n}_{Q_j} +\omega_0
\sum_{j} B^{\dagger}_{Q_j} B^{}_{Q_j}
\end{equation}
with
\begin{eqnarray}
%
%
B_{Q_j}^{\dagger} &=& {\bf U_{j,i}} b^{\dagger}_i \; , \; 
B_{Q_j}^{}={\bf U_{j,i}^{*}} b^{}_i ={\bf U_{-j,i}^{}} b^{}_i\,, \\
n_{Q_j}                 &=& \sum_{i}{\bf U_{j,i}} n_{i}\,,
\end{eqnarray}
where ${\bf U_{j,i}}= \frac{1}{\sqrt{N}} \exp{\{i Q_j R_i\}}$. 
$Q_j$ ($R_i$) denote the allowed 
momentum (translation) vectors of the lattice.
The phononic $Q=0$ mode couples to 
$n_{0}=\frac{N_{el}}{\sqrt{N}}$ which is a constant 
if working in a subspace with fixed number of electrons.
Thus the Hamiltonian decomposes into
${\cal H}=\tilde{\cal H}+{\cal H}_{Q=0}$, with 
\begin{equation}
{\cal H}_{Q=0}=- \sqrt{\varepsilon_p\omega_0}\,
({B^{\dagger}_{0}}+{B^{}_{0}})n_{0} + \omega_0  B^{\dagger}_{0} B^{}_{0}\,.
\end{equation}
Since $[\tilde{\cal H} \; , \; {\cal H}_{Q=0}]=0$ holds the eigenspectrum 
of ${\cal H}$ can be built up by the analytic solution for ${\cal H}_{Q=0}$ 
and the numerical results for $\tilde{\cal H}$.
Using the unitary transformation 
\begin{equation}
\label{UTransS}
\mbox{{\bf S}} (N_{el}) =  \exp{ \left\{ - \frac{N_{el}}{\sqrt{N}}
\sqrt{\frac{\varepsilon_p}{\omega_0}}
  ({B^{\dagger}_{0}}-{B^{}_{0}}) \right\} }\,,
\end{equation}
and introducing a shift of the phonon operators 
($B^{}_{0} \rightarrow B^{}_{0}+ \frac{N_{el}}{\sqrt{N}}
\sqrt{\frac{\varepsilon_p}{\omega_0}}$), 
we easily find the diagonal form of ${\cal H}_{Q=0}$
\begin{equation}
\bar{\cal H}_{Q=0} = \omega_0  B^{\dagger}_{0} B^{}_{0} 
- \varepsilon_p\frac{N^2_{el}}{N}\,.
\end{equation}
It represents a harmonic oscillator with 
eigenvalues and eigenvectors: 
\begin{eqnarray}
E_{\bar{l}} &=& 
\omega_0 \bar{l} -\varepsilon_p \frac{N^2_{el}}{N}\,,\\
| \bar{l} \rangle &=& \frac{1}{\sqrt{ \bar{l} !} }
  (B^{\dagger}_{0} )^{\bar{l}} | 0 \rangle\,.
\end{eqnarray}
The corresponding eigenenergies and  eigenvectors of 
${\cal H}_{Q=0}$ are $E_{l}=E_{\bar{l}}$ and 
\begin{equation}
\label{ezhsp}
|l, N_{el} ) =  \mbox{{\bf S}}^{\dagger} (N_{el}) | \bar{l}
  \rangle\,,
\end{equation}
respectively. That is, in the eigenstates of the Holstein model
a homogeneous lattice distortion occurs.
Note that the homogeneous lattice distortions are different 
in subspaces with different electron number. Thus excitations due to 
lattice relaxation processes show up in the one-particle spectral function.
Finally, eigenvectors and eigenenergies of ${\cal H}$ can be constructed by  
combining the above analytical result with the numerical determined 
eigensystem ($\tilde{E}_{n}^{(N_{el})}$; $|\tilde{\psi}_{n}^{(N_{el})} \rangle$) of $\tilde{\cal H}$:
\begin{eqnarray}
\label{defpsi}
E_{n,l}^{(N_{el})}             &=& \tilde{E}_{n}^{(N_{el})} + \omega_0 l 
- \varepsilon_p \frac{N^2_{el}}{N}\,,\\
|\psi_{n,l}^{(N_{el})} \rangle &=& |\tilde{\psi}_{n}^{(N_{el})} \rangle \otimes |l , N_{el} )\,.
\end{eqnarray}

The direct product notation for the eigenvectors of ${\cal H}$ 
implies that also the one-particle spectral functions, e.g. the 
single-particle spectral function 
\begin{eqnarray}
\label{defipe}
A_{K}^+(\omega)&=& \sum_{n,l} 
|\langle \psi_{n,l}^{(N_{el}+1)}|c_{K}^{\dagger} 
|\psi_{0,0}^{(N_{el})}\rangle|^2\;\delta (\omega-[E_{n,l}^{(N_{el}+1)}-E_{0,0}^{(N_{el})}])\,,
\end{eqnarray}
can also be decomposed in a numerical and an analytical contribution:
\begin{eqnarray}
\label{eqnuman}
{A}_{K}^+(\omega)&=&\sum_{l}
\rho(l,N_{el}+1, N_{el}) \;
\tilde{A}_{K}^+(\omega-[\omega_0 l- \varepsilon_p
(2 N_{el}+1)/N])\,. 
\end{eqnarray}
Using ED in combination with kernel polynomial moment expansion and 
maximum entropy methods~\cite{BWF98,RNS1,RNS2}, we first compute numerically 
the one-particle spectral function for $\tilde{\cal H}$,
\begin{eqnarray}
\tilde{A}_{K}^+(\tilde{\omega})&= &\sum_{n} |\langle \tilde{\psi}_{n}^{(N_{el}+1)}|c_{K}^{\dagger} 
|\tilde{\psi}_{0}^{(N_{el})}\rangle|^2\;\delta
(\tilde{\omega}-[\tilde{E}_{n}^{(N_{el}+1)}-\tilde{E}_{0}^{(N_{el})}])\,,
\end{eqnarray}
which does not include the effects of the $Q=0$ phonon mode.
In a second step, the final spectrum is constructed from Eq.~(\ref{eqnuman})  
by shifting $\tilde{A}_{K}^+(\tilde{\omega})$ by multiples of the bare 
phonon frequency and using the weight factors~\cite{Ro97}
\begin{equation}
\label{defrho}
\rho(l^{\prime},N_{el}^{\prime},N_{el})
                =|(N_{el}^{\prime},l^{\prime} | 0 , N_{el} )|^2 
                = \frac{x^{l^{\prime}}}{l^{\prime} !} \; \mbox{e}^{-x}\,,
\end{equation}
where 
$x=\frac{1}{{N}}\frac{\varepsilon_p}{\omega_0}(N_{el}^{\prime}-N_{el}^{})^2$. 
The $\rho(l^{\prime},N_{el}^{\prime},N_{el})$ quantify the 
relaxation process of the lattice distortion discussed above. 
Thus, each excitation in $\tilde{A}_{K}^+(\tilde{\omega})$ 
splits up into a band of peaks separated 
by $\omega_0$ which contains the total weight 
of the original excitation [$\sum_{l} \rho(l,N_{el}+1, N_{el}) =1$].

Fig.~\ref{EDexample} shows the contribution 
of the $Q=0$ phonon mode 
to $A_{K}^{\dagger} (\omega)$.  
The $Q=0$ excitations are separated 
by $\omega_0$ and interfere at higher energies with excitations 
forming broad bands. In Fig.\ref{EDexample}, e.g., 
five replications of the lowest peak are visible 
(with decreasing height and weight).
The integrated spectral density reflects the 
redistribution of spectral weight to higher energies but 
conserves the spectral weight when integrating the $Q\ne0$ peaks together 
with the corresponding $Q=0$ side bands.
Note that this behavior is observed in a finite system only,
i.e.,  $\rho(l^{\prime} \ge 1 ,N_{el} \pm 1,N_{el})$ vanishes 
in the thermodynamic limit $N\rightarrow \infty$.

In our finite cluster diagonalization, however, 
the separation of the phononic $Q=0$ mode significantly 
reduces the computational requirements.  
First, the dimension of the matrix to be diagonalized is 
reduced by a factor of $1+\frac{M}{N}$ because only $N-1$ 
instead of $N$ independent modes have to be considered.
Second, the $Q=0$ mode takes into account at least 
$\langle B_0^{\dagger}B_0 \rangle = \frac{\varepsilon_p}{\omega_0} 
\frac{N_{el}^2}{N}$ phonons in the ground state already.
Thus, the maximum number of phonons allowed per state, $M$, can be 
chosen much smaller for $\tilde{\cal H}$ than for the full Hamiltonian 
$ {\cal H}$ in order to achieve the same level of convergence.
The relevance of the separation of the phononic $Q=0$ mode for our 
numerical work becomes evident if one compares the dimension 
of the phononic Hilbert 
space used for the calculation in Fig.~\ref{EDexample} 
($N_{el}=4,\,N-1=7,\,M=28,\, \varepsilon_p/\omega_0=6$) 
$\tilde{D}_{ph}=6.7 \times 10^6$ with 
that required for the full Hamiltonian 
($N=8,\,M=28+12$) $D_{ph}=3.77 \times 10^8$, which 
is about two orders of magnitude larger.

\section{Peierls transition in the spinless fermion Holstein model} 
\label{Conc}

In this section we apply the PRM and ED techniques outlined in Secs.~II
and~III to the investigation of the metal insulator transition in the 
Holstein model. While both approaches are valid for any dimension
in principle, we restrict ourselves to the one-dimensional case and
spinless fermions, mainly because of the serious memory restrictions
within the ED calculations. On the other hand, focusing on the 
half-filled band case $N_{el}=N/2$, we know that the Peierls
instability, we are interested in, 
is most pronounced in low-dimensional systems.  
Throughout the numerical work we use the bare tight-binding  
electron dispersion $\varepsilon_{k} = -2t \cos{ka}$ (lattice constant $a=1$), 
and an Einstein phonon frequency $\omega_0/t = 0.1$. 
The Fermi level is defined 
to be the energy zero level and the temperature is $T=0$. 
In the following we shall vary the dimensionless electron-coupling constant 
$g/t$ or, equivalently, $\varepsilon_p/t$ [where 
due to $g =\sqrt{\varepsilon_p\omega_0}$ we have
$\varepsilon_p/t= (g/t)^2 (t/\omega_0)$].

\subsection{ED results}

In order to investigate lattice dynamical effects on the Peierls 
transition we first analyze the spectral density of 
single-particle excitations associated with the 
injection of an electron with wave number 
$K$, $A^+_K(\omega)$ (IPE), given by Eq.~(\ref{defipe}),
and the corresponding quantity for the emission of an
electron, $A^-_K(\omega)$ (PE). 
In $A^-_K(\omega)$ the destruction operator
$c_K$ connects the ground state of the Holstein model with $N_{el}$ electrons 
($|\psi_{0,0}^{(N_{el})}\rangle$) to all 
excited states of the system with $N_{el}-1$ electrons
($|\psi_{n,l}^{(N_{el}-1)}\rangle$). Of course, 
the {\it exact} determination of $A^{\pm}_K(\omega)$ 
for coupled electron-phonon
systems is an tremendous numerical task, which requires the repeated
solution of eigenvalue problems with dimensions of the order of
$10^{10}$. At present this can only be achieved by 
employing elaborate numerical techniques, e.g. 
the kernel polynomial expansion method, on leading edge 
supercomputers.\cite{Num98}

Let us first consider the adiabatic ($\alpha=0.1$) 
weak coupling regime $\varepsilon_p/t\ll 1$.
The data presented for the single-particle spectral
function in Fig.~\ref{edak01} give clear evidence 
that the system behaves like a metal. The Fermi energy
($K_F=\pi/2$) is located in the center of an only weakly renormalized
band (the bandwidth is about 4t, which is the value 
for noninteracting system). The band dispersion can 
be derived by tracing the lowest (uppermost) peak in 
each $K$ sector of the IPE (PE) spectra. 
Most notably these peaks have a spectral weight close
to one. As an effect of the electron-phonon coupling
phonon satellites separated by the bare phonon 
frequency $\omega_0/t$ occur in the vicinity of 
the tight-binding band electron levels. These excitations, 
however, have extremely small (electronic) spectral weight.

If we increase the electron-phonon interaction a gap
feature starts to develop in the single-particle spectra 
at the Fermi momentum. This is the situation shown 
in Fig.~\ref{edak06}. Obviously a critical interaction 
strength is necessary to trigger the Peierls transition
at finite phonon frequencies. This has to be contrasted to 
the result obtained for the adiabatic limit $\omega_0=0$, 
where a Peierls instability occurs at any finite electron-phonon coupling. 
Figure~\ref{edak06} also demonstrates the mixed (electron-phonon)
nature of the excitations. Now the spectral weight is almost 
equally distributed among the different peaks in each $K$ sector. 

Finally we examine the behavior of the single-particle spectra 
in the strong electron-phonon coupling regime. As can be seen from 
Fig.~\ref{edak16}, a wide band gap emerges, indicating
massive charge excitations accompanied by multi-phonon absorption 
and emission processes. The Peierls distortion of the lattice   
is directly connected to a charge-density-wave formation 
(we found almost localized electrons at every second site).
As a result a symmetry-broken ground state may occur in the infinite system,
reflecting true long-range order. Now the Fermi energy is located at the
center of the band gap and consequently the system shows insulating behavior. 

To supplement the discussion we display in Fig.~\ref{phdis}
the contribution of states with $m$ phonons contained in the ground state.  
From top to the bottom the coupling strength increases.
At weak electron-phonon coupling  $\varepsilon_p/t=0.1$
the zero-phonon state is clearly the dominant one.
States with more than two phonons are negligible.
For the Peierls phase ($\varepsilon_p/t = 1.6$) the phonon distribution of 
the ground state follows a Poisson-like distribution, where the 
ratio $\varepsilon_p/\omega_0= 16$ gives a good estimate of the
mean phonon number.

\subsection{PRM results}

In order to integrate 
the renormalization equations derived in Sec.~II we consider 
a lattice of $N=1000$ sites in one dimension. 
The width of the energy shell $\Delta \lambda$ was taken to be
somewhat smaller than the typical smallest energy spacing of 
the eigenstates of ${\cal H}_{0,\lambda}$.    
Fig.~\ref{fig1}  shows  the electronic spectral functions 
$A_k^+(\omega)$ and $A_k^-(\omega)$ [calculated from Eqs.~\eqref{35} and
\eqref{36}], and the renormalized phonon dispersion $\tilde{\omega}_{q}$
from the PRM approach for a  $g$-value of  $g/t=0.10$ 
($\varepsilon_p/t =0.1$) which is below the critical electron-phonon coupling
$g_c$. The wave number $k$ is fixed  
to $k_F= \pi/2$. Note that for $g$ values smaller than $g_c$ the 
system is in a Luttinger liquid (metallic) phase. As is seen from 
Fig.~\ref{fig1} the one-particle spectrum consists of a main 
pole and of satellites due to the coupling of the electrons
to phonons. They follow from the first 
and to the second and third terms on the r.h.s.~of  
Eqs.~\eqref{36} and \eqref{37} and will be denoted in the following as 
coherent and incoherent excitations. Note that due to the used
approximations to derive the PRM equations it is beyond the 
scope of the present approach to discuss possibly differences 
between a Luttinger liquid and a Fermi liquid  behavior in the one-particle
spectral functions.

As was mentioned before we restrict ourselves to $k= k_F$.  Therefore  the 
coherent excitation is located at the Fermi level 
[dotted line in Fig.~\ref{fig1}\;(a)]. It
is the dominant excitation of the spectrum since its  pole
strength $\tilde{\alpha}_{k_F}^2$  is 
close to its maximum value of $1$. Due to 
(\ref{38}) the intensity sum over all coherent and 
incoherent excitations  is equal to $1$. The 
system is in a metallic state. Note the small finite 
onset of the incoherent parts in $A_k^+(\omega)$ and $A_k^-(\omega)$ close  
to the Fermi level which is due to the finite phonon frequency 
$\tilde{\omega_q} \approx \omega_0$ (for small $g$-values). 
In Fig.~\ref{fig1}\;(b) the renormalized phonon frequency 
$\tilde{\omega}_q$ is shown as a function of the wave number.
Due to the coupling between the 
phononic and the electronic degrees of freedom, 
$\tilde{\omega}_q$ has gained some dispersion.   
For the chosen $g$ value this dispersion is 
relatively small except for  wave numbers $q$ close to the 
Brillouin-zone center and close to the
Brillouin-zone boundary ($k_0= \pm \pi$).
This feature can be understood from the phonon renormalization 
equation (\ref{24}). There, the second order renormalization 
contribution corresponds  to a frequency dependent energy  
shift which is similar to that known from the 
phonon self-energy contribution of second order. 
The dispersion results from particle-hole excitations 
$c_{k}^\dagger c_{k+q}$ 
around the Fermi level with a wave number transfer $q$
which is either small ($q \approx 0$) or about 
$q \approx k_0=\pm \pi$ (since at half-filling
$\varepsilon_{k_F} \approx \varepsilon_{k_F + k_0}$). Note that 
Fig.~\ref{fig1} shows an overall agreement with the ED 
results of Fig.~\ref{edak01} where the same parameter have been used. 

\bigskip
In Fig.~\ref{fig2}\; a) and b) the same quantities are shown for 
$g/t= 0.30$ \; ($\varepsilon_p/t = 0.9$). In this case the coupling strength 
is somewhat below the critical value $g_c$. 
This can be seen from the pole strength $\tilde{\alpha}_{k_F}^2$ of the 
coherent pole which has considerably decreased from almost $1$  
to a value already very small compared to $1$. 
Since the vanishing of $\tilde{\alpha}_{k_F}^2$ has to be interpreted as a 
signature of the metal-insulator transition the system is still in the  
metallic state.
In Fig.~\ref{fig2}\;b) the renormalized 
phonon frequency $\tilde{\omega}_q$ is shown. For large 
wave numbers approaching the Brillouin-zone boundary 
a strong softening is observed. 
Only for wavevectors extremely close 
to the boundary a sudden stiffening is  found again. This  feature  is 
reminiscent of the behavior at $k_0$ discussed before.  

\bigskip
This stiffening is absent in Fig.~\ref{fig3}\;b) where the PRM result
for the renormalized phonon frequency is shown for $g/t=0.34$ \;
($\varepsilon_p/t =1.156$) which is somewhat larger than the 
critical $g$ value (within the PRM approach).
In this case the phonon excitation becomes negative very close to the 
Brillouin-zone boundary $k_0$. The 
negative phonon frequency indicates the occurrence of an instability at the
critical coupling strength $g_c$ 
which is associated with the transition of the system to a 
Peierls state. Note that  for $g>g_c$ the present PRM
treatment breaks down since a 
possible shift  of the ionic equilibrium positions was not taken into 
account. A discussion of the phonon dynamics in the Peierls 
state will be subject to further research. 
However, in contrast to  the discussion of the 
phonon modes the discussion of the 
electronic excitations within the PRM method is not restricted to $g$ values 
$g <g_c$ but can be extended to some small range above $g_c$ (see below).
In Fig.~\ref{fig3}\; a) the  spectral functions $A_k^+(\omega)$
and $A_k^-(\omega)$ from the PRM treatment are shown for 
the same $g$ value $g/t=0.34$. The spectrum has changed from a 
strongly peaked distribution for smaller $g$ to a more Poisson-like 
shape. The coherent pole has completely vanished 
$\tilde{\alpha}_{k_F}^2 =0$ and a gap has opened in the 
spectrum at the Fermi level. The system has undergone a phase 
transition to a nonmetallic state.  To verify that 
the gap does not depend on the lattice size we have applied the PRM 
to a rather large system with 1000 sites but 
also to smaller systems with 800, 500, 200, and 100 sites. In all 
cases the size of the gap was practically independent of the 
system size so that the gap can be considered as an intrinsic 
property of the Holstein model. As in any renormalization 
group procedure the results could depend on the actual cluster size
due to additional degeneracies imposed by symmetry, especially for small 
systems. Such a dependence was not found for the systems considered 
here.

\bigskip
To find the critical coupling strength at which the phase transition occurs 
we have plotted the 
coherent pole strength $\tilde{\alpha}_{k_F}^2$ at 
wave vector $k=k_F$ for different values of $g$ (Fig.~\ref{fig4}).  Note the 
strong decrease of $\tilde{\alpha}_{k_F}^2$ with 
increasing $g$. A closer inspection of the data shows that  
at $g=g_c \approx 0.31 t$ the pole strength 
becomes  zero which marks the transition of the 
system to the nonmetallic state. This value is somewhat 
larger than the critical value found from the exact diagonalization 
[$g_c/t \approx 0.24$ ($\varepsilon_{p}/t\approx0.6$), 
compare Fig.~\ref{edak06}]. 
Note, however, that  
in the PRM approach a rather large system with 1000 lattice sites was used, 
whereas in the  ED the system had to be restricted to 8 sites. 
For the smaller systems with $800, 500, 300, 200$, and $100$ sites the  
critical coupling was slightly  smaller and approximately
$g_c/t  \approx 0.30$. A comparison with the critical value $g_c/t 
\approx 0.28$, obtained from DMRG calculations of 
Refs.~\onlinecite{BMH98,FWH04},
shows that the critical values from PRM   might be somewhat too large.  
The difference is probably due to the 
factorization approximation in the PRM approach and to the 
simplified ansatz \eqref{15} for the operator form of the generator 
$X_{\lambda,\Delta \lambda}$.  

An alternative criterion to determine the critical coupling strength
is to take that value at which the renormalized phonon frequency 
$\tilde{\omega}_q$ at the Brillouin-zone boundary $q= \pm k_0$
vanishes. In this way one finds for the system with 
1000 sites $g_c/t \approx 0.30$, i.e., a result 
which is somewhat smaller than that found from the vanishing of 
$\tilde{\alpha}_{k_F}^2$. Note however
that the latter $g_c$ value seems to be the more reliable one. 
This can be understood from the comparison of the renormalization equations    
\eqref{32} for $\alpha_{k,\lambda}$ 
and \eqref{24} for $\omega_{q,\lambda}$. As can be seen, 
for $\alpha_{k,\lambda}$ an approximate exponential decay 
with $\lambda$ is found whereas a logarithmic decay follows for 
$\omega_{q,\lambda}$. Thus, since the 
dominant renormalization contributions always occur at small $\lambda$,
the numerical errors are much smaller for $\alpha_{k,\lambda}$ 
than for $\omega_{q,\lambda}$. Therefore, within 
the present renormalization approach the 
critical coupling strength found from the vanishing of the 
coherence strength $\tilde{\alpha}_{k_F}^2$ is probably 
more reliable than that given by the vanishing 
of $\tilde{\omega}_{q=k_0}$. On the other hand, from comparison 
with other approaches one has to admit that the value of $g_c/t \approx
0.31$ is possibly somewhat too large probably due to the 
factorization approximation which was used to derive 
the renormalization equations.

\bigskip
As was mentioned above,  the PRM
approach is only valid in the metallic regime $g<g_c$. However, one 
can assure oneself of the fact  that it may also 
be applied in a small parameter regime $g \geq g_c$: We again
consider ${k}={k}_F$. Due to the $\Theta$-functions 
$\Theta_{{k},{q}}$ in all renormalization equations a 
renormalization approximately occurs when the energy difference 
$|\omega_{{q},\lambda} + \varepsilon_{{k}, \lambda} 
- \varepsilon_{{k}+{q},\lambda}|$ lies within a small energy shell
between $\lambda$ and $\lambda - \Delta \lambda$. Moreover, the most 
dominant renormalization processes take place for small 
cutoff $\lambda$.  Therefore,
$\varepsilon_{{k}_F + {q},\lambda} \approx 
\varepsilon_{{k}_F, \lambda}$ has to be fulfilled (where a small 
phonon frequency was assumed).  It follows that  
${q} \approx \pm {k}_0$, where ${k}_0$ is the zone-boundary
wave vector.  According to (\ref{32}) the coherent pole strength 
$\alpha_{{k}_F,\lambda}^2$ will mostly be renormalized 
by wave numbers ${q} \approx {k}_0$. In contrast 
$\beta_{{k}_F, {k}_0, \lambda}$ and                 
$\gamma_{{k}_F, {k}_0, \lambda}$ are both renormalized only
once when $\lambda$ is small. According to (\ref{32}) 
$\alpha_{{k}_F, \lambda}$ 
becomes zero for small $\lambda$ in the insulating regime $g>g_c$, so that 
the renormalization of $\beta_{{k}_F, {k}_0, \lambda}$ and 
$\gamma_{{k}_F, {k}_0, \lambda}$  is negligible due to (\ref{33})
and (\ref{34}). Note that the incoherent excitations in 
(\ref{35}) and (\ref{36}) close to ${q} \approx 
{k}_0 $ contain negative phonon energies $\tilde{\omega}_{q}$. 
However, they do not influence the electronic spectral functions.

\section{Summary}

In this paper the electron-phonon coupling in the one-dimensional Holstein
model at half-filling has been investigated using both the projector-based 
renormalization method (PRM) and an refined exact diagonalization technique 
in combination with the kernel polynomial method (ED). In this system a
metal-insulator transition occurs, accompanied by the formation of Peierls
distorted state. This transition has been analyzed in terms of the (inverse)
photoemission spectral functions, where the phonon dynamics is
fully  taken into account.

Different from the ED the present PRM treatment is 
restricted to values $g$ smaller or equal to the 
critical electron-phonon coupling $g_c$.  The discussion of the coupling regime
$g$ larger than $g_c$ is more complicated and has to be 
postponed to a future investigation. However, it turns out that
the results for the electronic spectral functions  are not restricted to
$g< g_c$ but are also valid in a small regime regime 
$g$ above $g_c$. Therefore, the opening of an electronic gap can be observed. 
The reason is that electronic properties are rather insensitive against a 
small number of instable phonon modes close to the Brillouin-zone boundary. 

Although both the PRM and the ED technique are valid for any dimension, we
have restricted ourselves to a one-dimensional lattice.
The (inverse) photoemission spectral functions from both 
approaches clearly indicate a metal-insulator transition 
and the opening of a gap if the electron-phonon coupling 
becomes larger than a finite critical value. The single-particle spectra 
excitations are accompanied by multi-phonon absorption and
emission processes in the Peierls phase.
The PRM provides a renormalized phonon dispersion 
which shows a softening at the Brillouin-zone boundary. 
This effect even occurs for coupling
strength $g$ smaller than the critical value $g_c$ so that the phonon
softening can be understood as a precursor effect of the gap formation.

\section*{Acknowledgements}

We appreciate valuable discussions with A.~R.~Bishop, 
K.~Meyer, T.~Sommer, and A.~Wei{\ss}e. 
Work in Dresden, Davis, Erlangen and Greifswald
was partially supported by the Deutsche Forschungsgemeinschaft 
through the research programs SFB 463,  under Grant No. HU 993/1-1, 
the Bavarian Competence Network for High Performance, and SPP 1073,
respectively. Special thanks go to the HLRN Berlin-Hannover, 
RRZ Erlangen, and LRZ M\"unchen
for the generous granting of their computer facilities. 

\appendix

\section{Transformation of the operators}\label{A}

In this appendix we evaluate the transformation from 
$\lambda$ to $(\lambda -\Delta \lambda)$. As an example let us consider
the operator $c_{{\bf k} }^\dagger c_{{\bf k}}$
\begin{eqnarray}
\label{A.1}
e^{X_{\lambda, \Delta \lambda}}  
c_{{\bf k} }^\dagger c_{{\bf k}}
e^{-X_{\lambda, \Delta \lambda}} 
&=&
e^{{\bf X}_{\lambda, \Delta \lambda}}  
\left(c_{{\bf k} }^\dagger c_{{\bf k}}\right)
= \sum_{n=0}^\infty \frac{1}{n!} {\bf X}_{\lambda,\Delta \lambda}^n
\left(   
c_{{\bf k} }^\dagger c_{{\bf k}}\right)
\end{eqnarray}
Here, a new super-operator ${\bf X}_{\lambda, \Delta \lambda}$
was introduced which is defined by the commutator of 
the generator $X_{\lambda, \Delta \lambda}$ with operators 
${\cal A}$ on which ${\bf X}_{\lambda, \Delta \lambda}$ 
is applied, ${\bf X}_{\lambda, \Delta \lambda}{\cal A}=
[X_{\lambda, \Delta \lambda}, {\cal A}]$. Thus we have to evaluate 
the  commutators on the r.h.s.~of (\ref{A.1})
\begin{eqnarray}
\label{A.2}
[X_{\lambda, \Delta \lambda}, 
c_{{\bf k} }^\dagger c_{{\bf k}}] &=&
\frac{1}{\sqrt N} \sum_{\bf q} \Big\{
B_{{\bf k}-{\bf q}, {\bf q},\lambda} 
\Theta_{{\bf k}-{\bf q}, {\bf q}}(\lambda,\Delta\lambda)
\left(b_{\bf q}^\dagger c_{{\bf k}-{\bf q}}^\dagger c_{{\bf k}}
+ b_{\bf q} c_{{\bf k}}^\dagger c_{{\bf k}-{\bf q}}
\right) \nonumber\\
&& \hspace*{2cm} -[{\bf k} \rightarrow ({\bf k}+{\bf q}) ]
\Big\}
\end{eqnarray}
By applying ${\bf X}_{\lambda, \Delta \lambda}$ twice on 
$c_{{\bf k} }^\dagger c_{{\bf k}}$ products composed of 
four fermionic and of two fermionic and two bosonic operators 
occur. In order to keep only operators which are also present in 
${\cal H}_{(\lambda- \Delta \lambda)}$ a factorization is performed. 
One obtains
\begin{eqnarray}
\label{A.3}
\lefteqn{
  \frac{1}{2} [X_{\lambda, \Delta \lambda}, [X_{\lambda, \Delta \lambda}, 
  c_{{\bf k} }^\dagger c_{{\bf k}}]] \,=\,
}&& \\
&=& 
\frac{1}{N} \sum_{\bf q} 
\left\{
  B^2_{{\bf k}-{\bf q},{\bf q},\lambda} 
  \Theta_{{\bf k}-{\bf q}, {\bf q}}(\lambda,\Delta\lambda) 
  (n_{{\bf k}\lambda}^c + n_{{\bf q},\lambda}^b ) \;
  c_{{\bf k}-{\bf q}}^\dagger c_{{\bf k}-{\bf q}}
\right.
\nonumber\\
&&
\left.
  \qquad -\,
  B^2_{{\bf k}-{\bf q}, {\bf q}, \lambda} 
  \Theta_{{\bf k}-{\bf q}, {\bf q}}(\lambda,\Delta\lambda)
  (1 - n_{{\bf k}-{\bf q}, \lambda}^c + n_{{\bf q},\lambda}^b ) \;
  c_{{\bf k}}^\dagger c_{{\bf k} }
\right.
\nonumber\\
&& 
\left.
  \qquad +\, B^2_{{\bf k}-{\bf q}, {\bf q},\lambda} 
  \Theta_{{\bf k} -{\bf q},{\bf q}}(\lambda,\Delta\lambda)
  (n^c_{{\bf k}-{\bf q}, \lambda} -n^c_{{\bf k} \lambda} ) \;
  b_{\bf q}^\dagger b_{\bf q} 
  \qquad + [{\bf k} \rightarrow {\bf k} +{\bf q}] \; 
\right. \Big\}
\nonumber
\end{eqnarray}
where the expectation values $n_{{\bf k},\lambda}^c$, $n_{{\bf q},\lambda}^b$
have been defined in (\ref{19}). The third order term 
of (\ref{A.1}) again gives 
interaction-type contributions. Thus, all operator terms 
appearing on the r.h.s.~of (\ref{A.1}) are traced back to those which also 
appear in ${\cal H}_\lambda$. This property enables us to evaluate all higher 
order commutators with $X_{\lambda, \Delta \lambda}$ and thus 
the transformation (\ref{A.1}). the result  is given in (\ref{18}).



\newpage

\begin{figure}[ht]
\caption{\label{EDexample} Low-energy excitations of the single-particle
spectral function  $A_{K}^{+} (\omega)$ 
and the corresponding integrated 
spectral density [$S_{K}^{+} (\omega)= 
\int_{-\infty}^{\omega} A_{K}^{+} (\omega^{\prime}) 
d \omega^{\prime}$] at $K=\pm \pi/2$ for $\varepsilon_p/t=0.6$ and 
$\omega_0/t=0.1$. Dashed (solid) lines show the spectrum including 
all phonon modes (without the $Q=0$ phonon mode). }
\end{figure}

\begin{figure}[h]
\caption{\label{edak01}(Color online) 
Wave-number-resolved spectral densities for photoemission 
[$A_{K}^-(\omega)$; dashed (red) lines] 
and inverse photoemission [$A_{K}^+(\omega)$; solid (black) lines]
in the metallic state ($\varepsilon_p/t \ll 1$).  
The corresponding integrated densities
$S_{K}^{\pm}(\omega)$ are given by bold lines.
All ED data were obtained for an eight-site system 
with periodic boundary conditions. Note that the sum rules
are fulfilled.}
\end{figure}

\begin{figure}[h]
\caption{\label{edak06}(Color online) PE [dashed (red) lines] and IPE 
[solid (black) lines] spectra
near the metal to Peierls insulator transition  point.}
\end{figure}

\newpage
\begin{figure}[h]
\caption{\label{edak16}(Color online) PE and IPE spectra in the Peierls phase.}
\end{figure}

\begin{figure}[h]
\caption{\label{phdis}(Color online) 
Phonon distribution in the ground state of the spinless Holstein
model for three characteristic coupling strengths.}
\end{figure}

\begin{figure}[h]
\caption{Results from the PRM approach for $g/t=0.10$ 
($\varepsilon_p/t =0.1$): 
(a) Electronic spectral functions 
$A_k^+(\omega)$(black) and $A_k^-(\omega)$(red) where the wave number $k$ is
fixed to $k_F=\pi/2$. The coherent
excitation peak (dotted line) 
is at $\omega=0$. (b) Renormalized  phonon dispersion
$\tilde{\omega}_{q}$. The original phonon frequency is $\omega_{0}/t = 0.1$.} 
\label{fig1}
\end{figure}

\newpage
\begin{figure}[h]
\caption{Same quantities as in Fig.~\ref{fig1}
  from the PRM approach for a $g$-value of $g/t = 0.30$ 
  ($\varepsilon_p/t = 0.9$) somewhat below
  the critical value $g_{c}/t \approx 0.31$ ($\varepsilon_p/t = 0.96$).}
\label{fig2}
\end{figure}

\begin{figure}[h]
\caption{Same quantities as in Fig.~\ref {fig1} from the PRM approach for
$g/t= 0.34$ \; ($\varepsilon_p/t = 1.16$) which is somewhat larger 
than $g_c/t \approx 0.31$ ($\varepsilon_p/t = 0.96$). 
Note that the coherent pole 
has vanished and a charge gap has opened (Fig.~8 a))}
\label{fig3}
\end{figure}

\begin{figure}[h]
\caption{Coherent pole strength $\tilde{\alpha}_{k_F}^{2}$ from 
the PRM approach as function of the electron-phonon coupling $g$  
for a system with 1000 lattice sites} 
\label{fig4}
\end{figure}


\clearpage
\begin{center}
  \bigskip
  \scalebox{1.0}{
    \includegraphics[70,500][526,742]{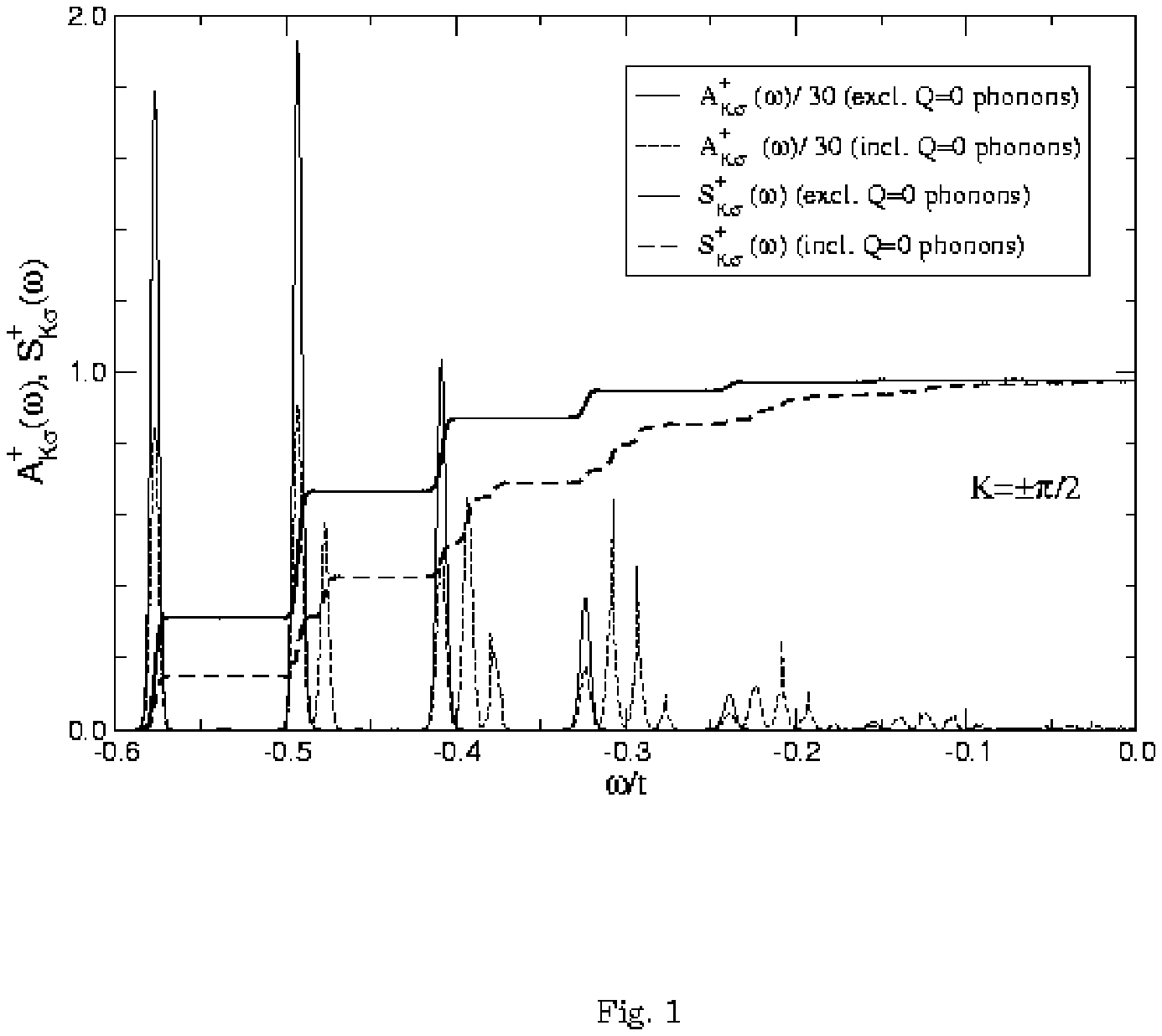}
  }
\end{center}

\clearpage
\begin{center}
  \bigskip
  \scalebox{1.0}{
    \includegraphics[70,500][526,742]{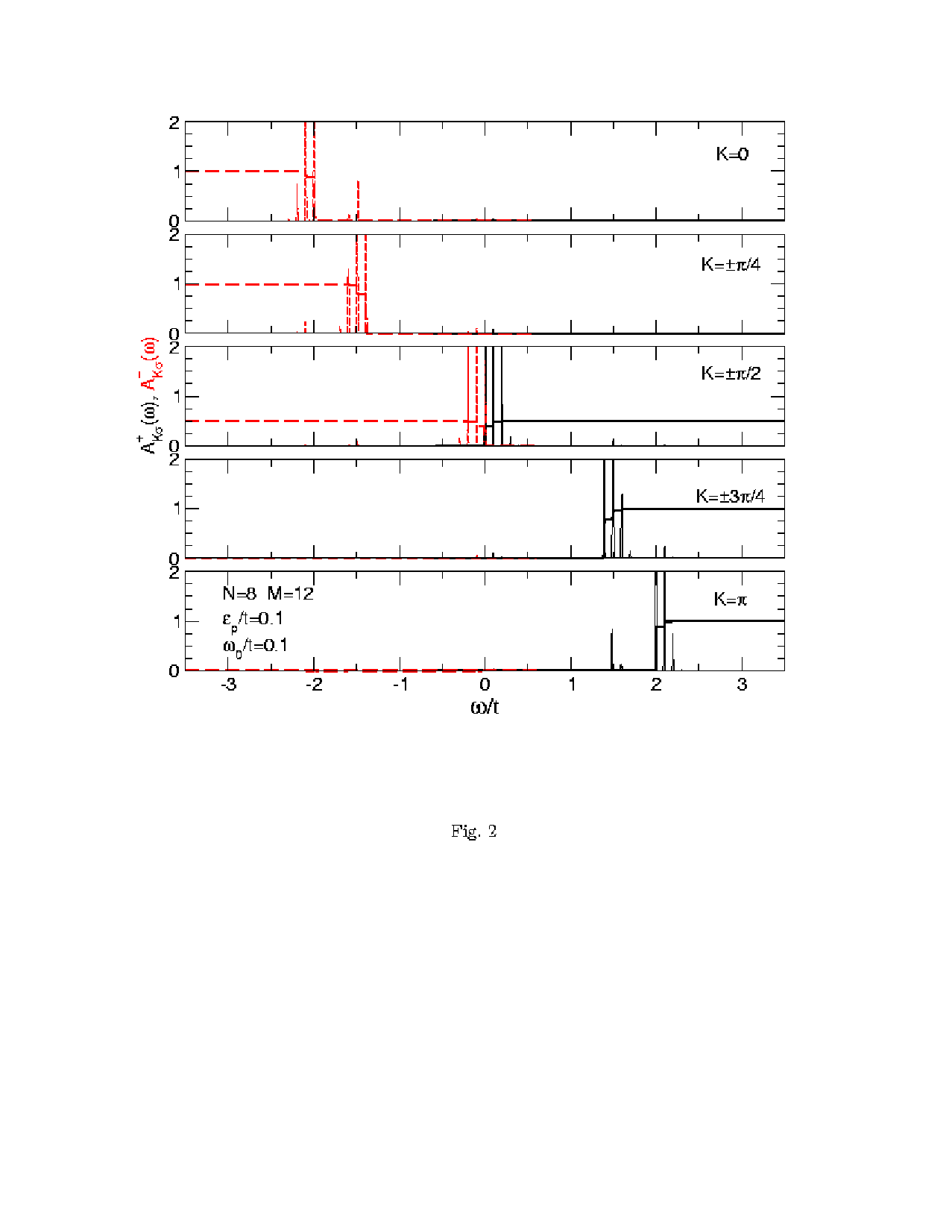}
  }
\end{center}

\clearpage
\begin{center}
  \bigskip
  \scalebox{1.0}{
    \includegraphics[70,500][526,742]{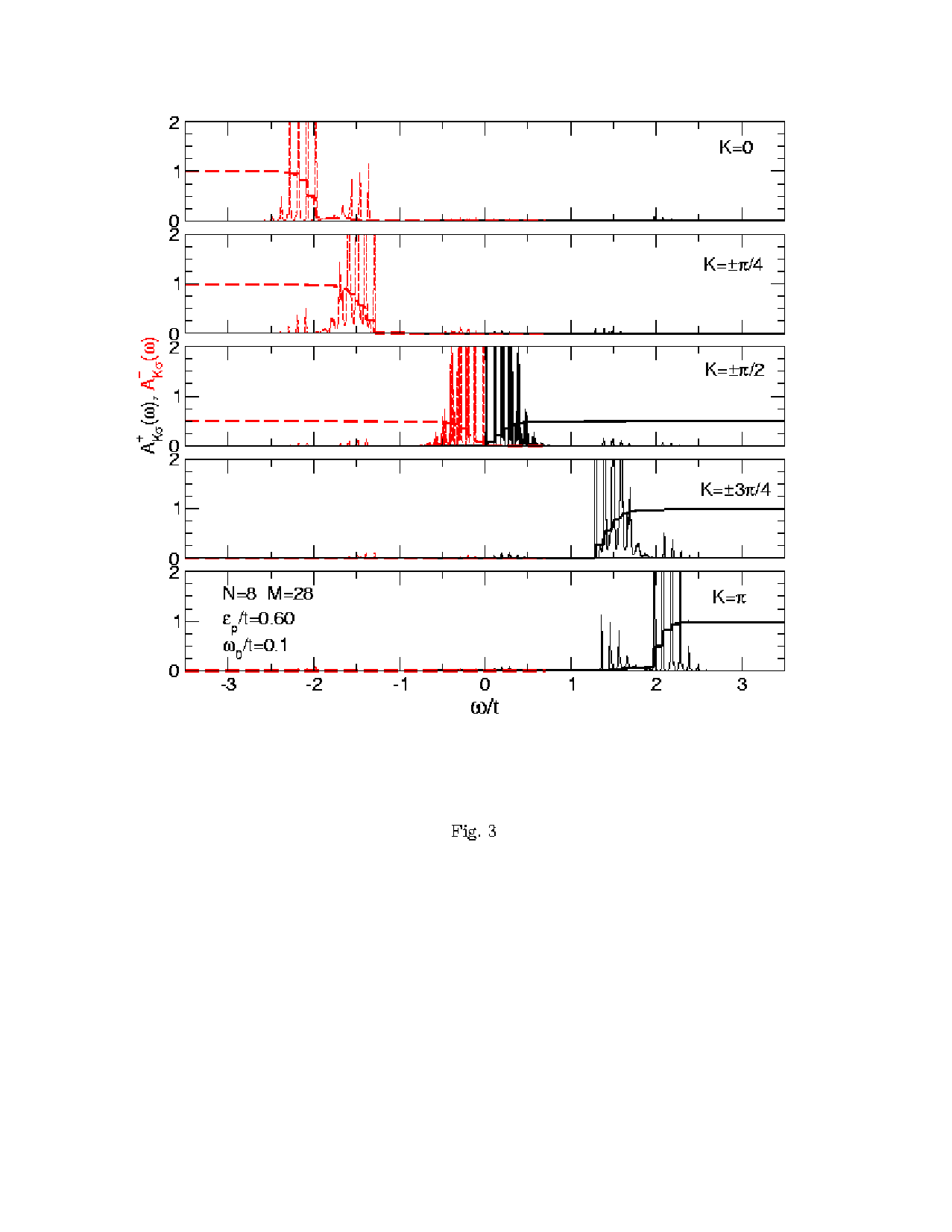}
  }
\end{center}

\clearpage
\begin{center}
  \bigskip
  \scalebox{1.0}{
    \includegraphics[70,500][526,742]{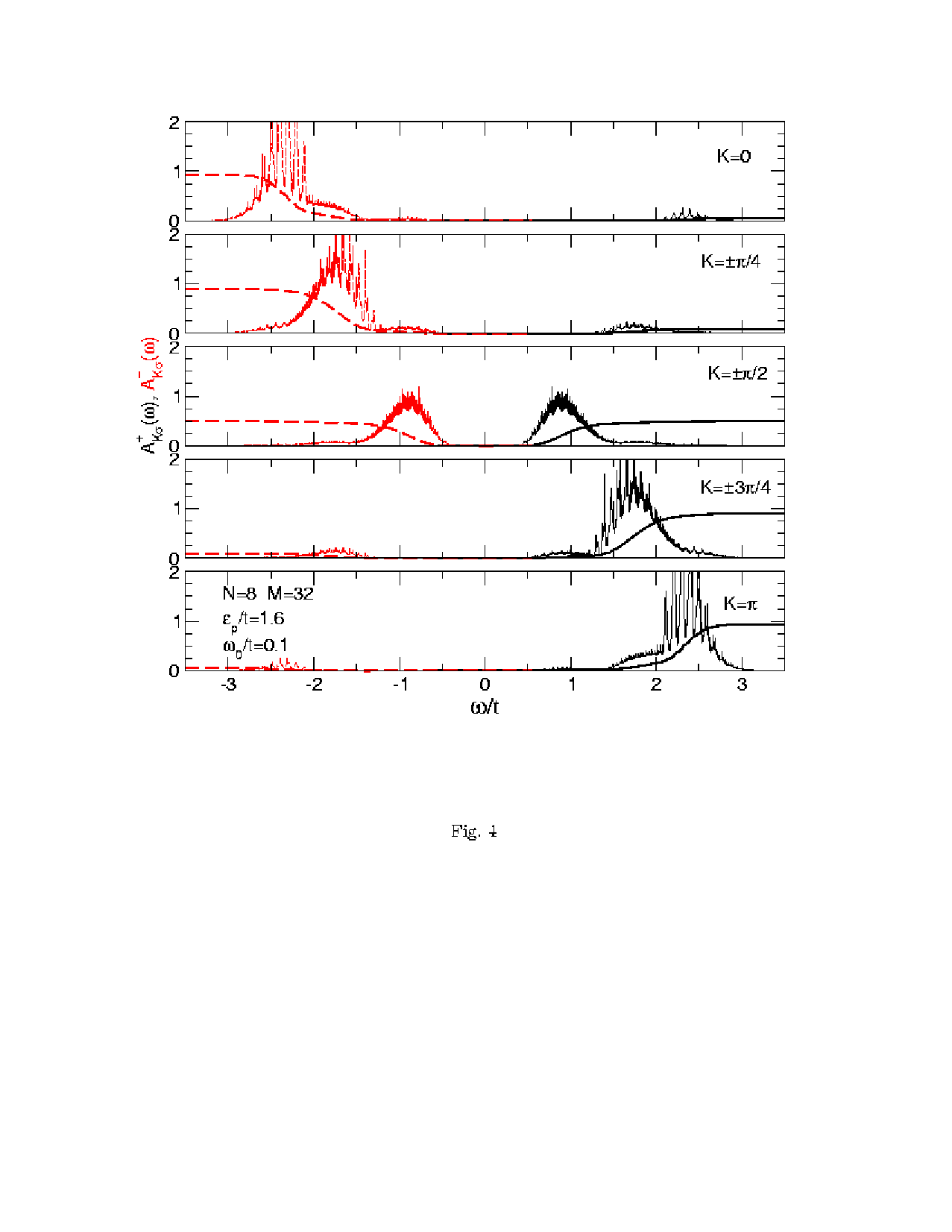}
  }
\end{center}

\clearpage
\begin{center}
  \bigskip
  \scalebox{1.0}{
    \includegraphics[70,500][526,742]{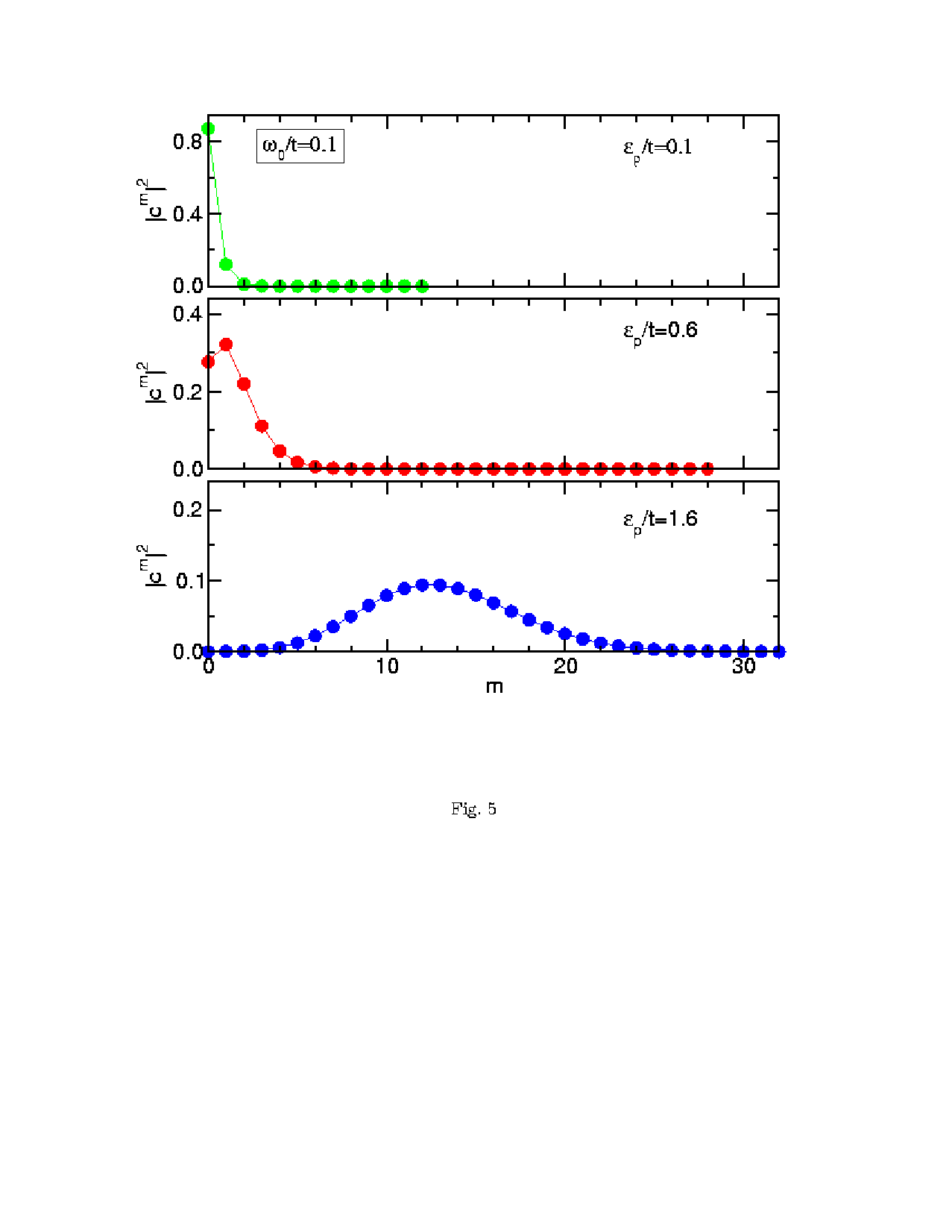}
  }
\end{center}

\clearpage
\begin{center}
  \bigskip
  \scalebox{1.0}{
    \includegraphics[70,500][526,742]{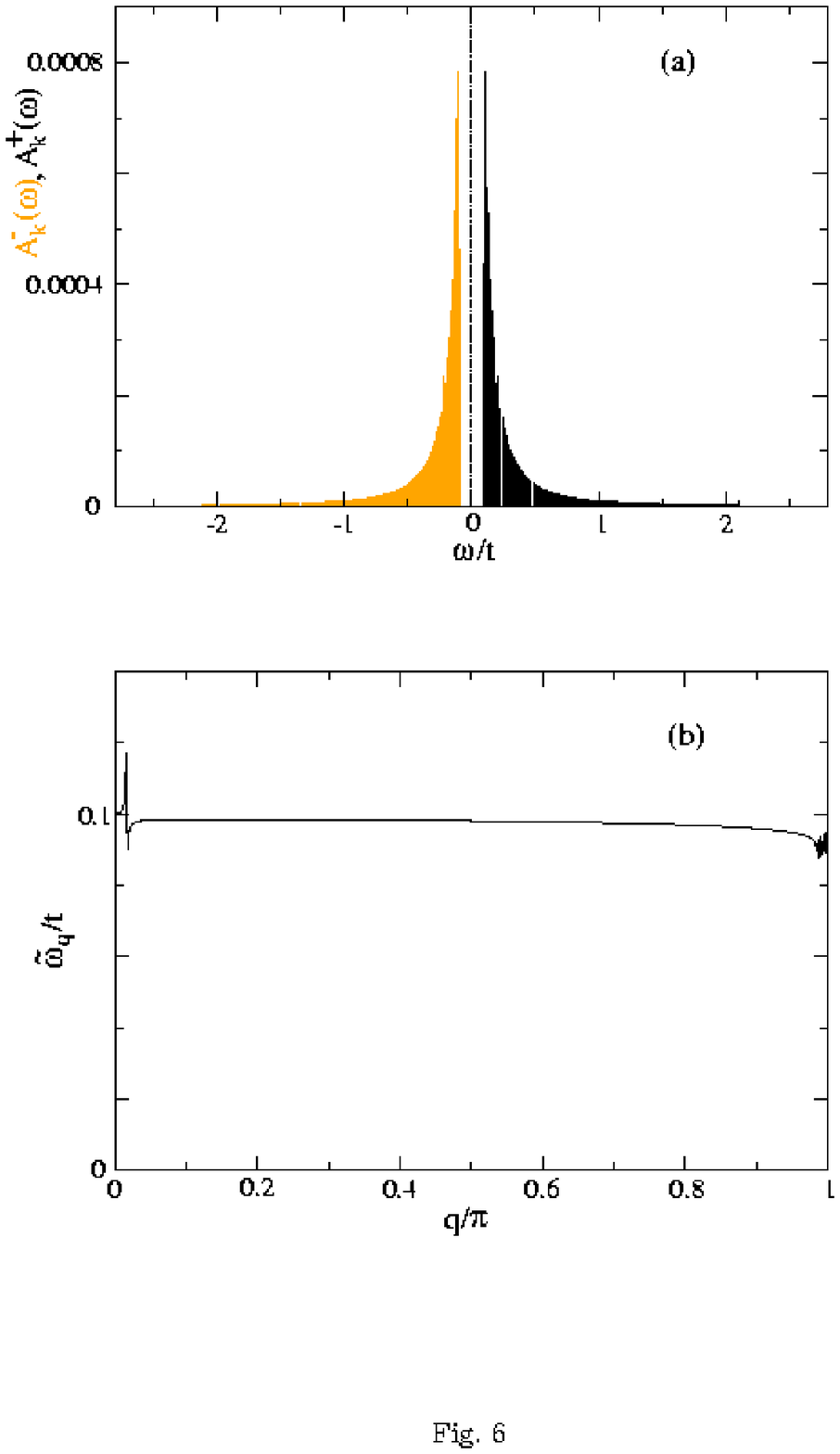}
  }
\end{center}

\clearpage
\begin{center}
  \bigskip
  \scalebox{1.0}{
    \includegraphics[70,500][526,742]{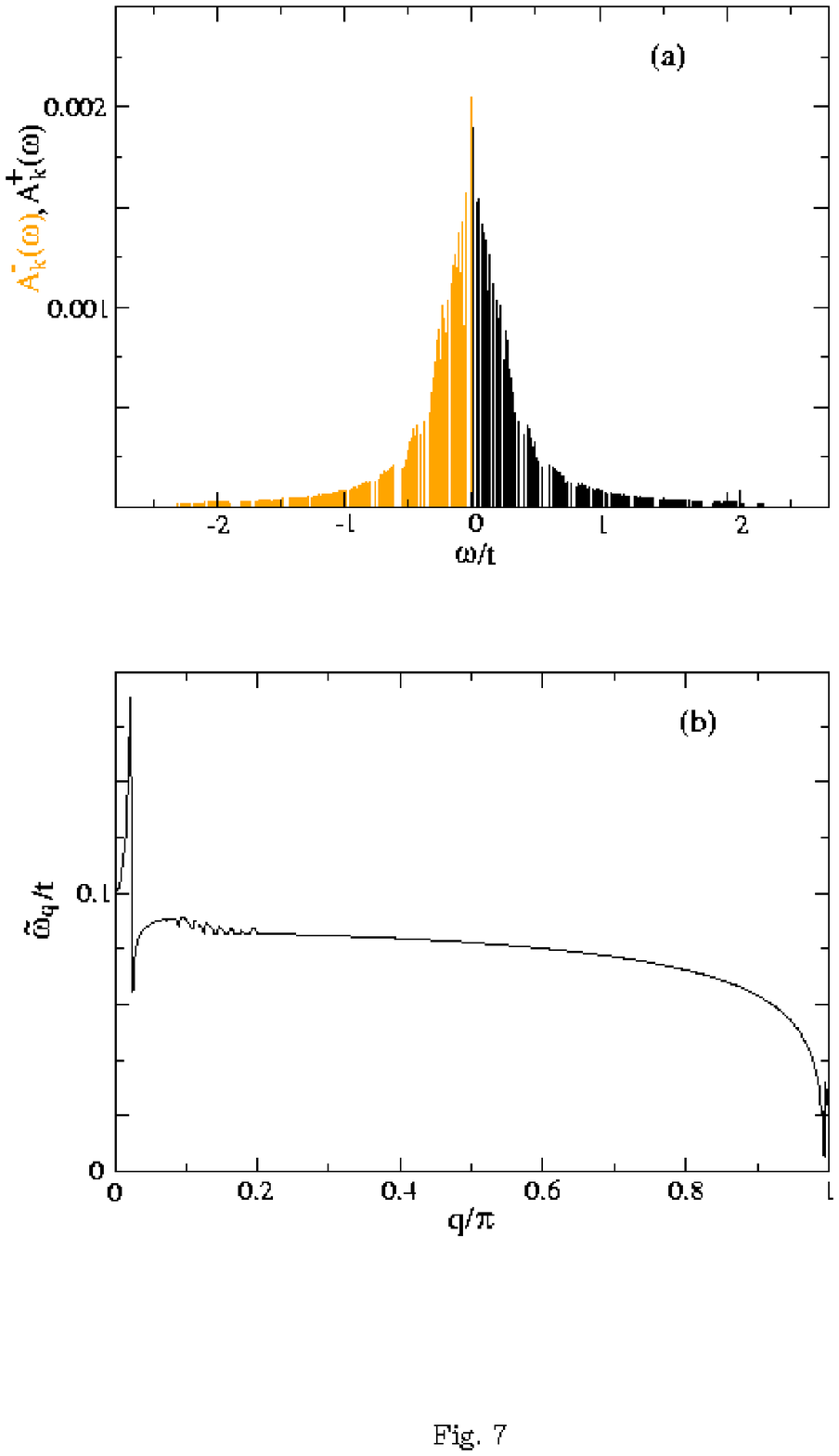}
  }
\end{center}

\clearpage
\begin{center}
  \bigskip
  \scalebox{1.0}{
    \includegraphics[70,500][526,742]{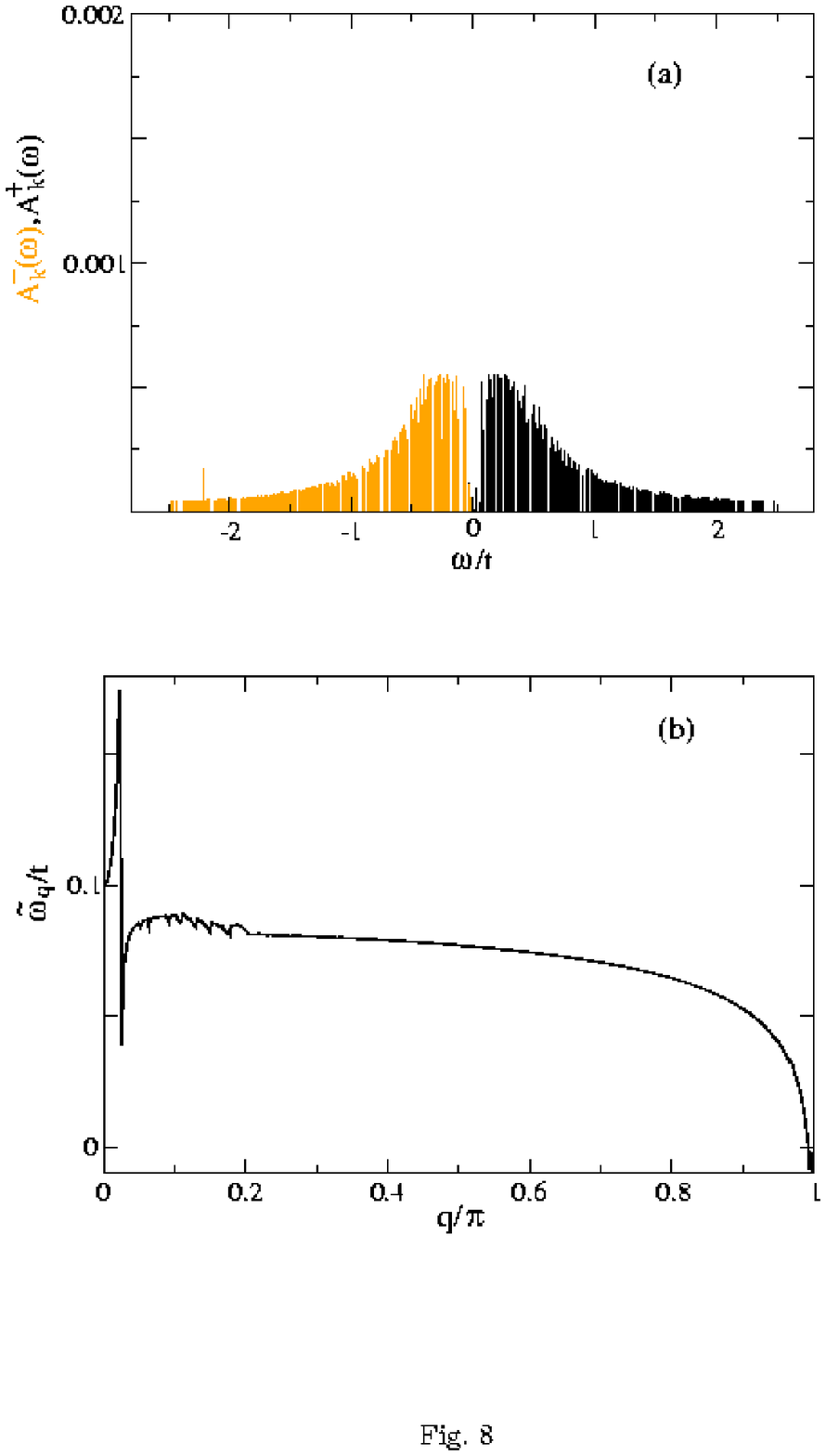}
  }
\end{center}

\clearpage
\begin{center}
  \bigskip
  \scalebox{1.0}{
    \includegraphics[70,500][526,742]{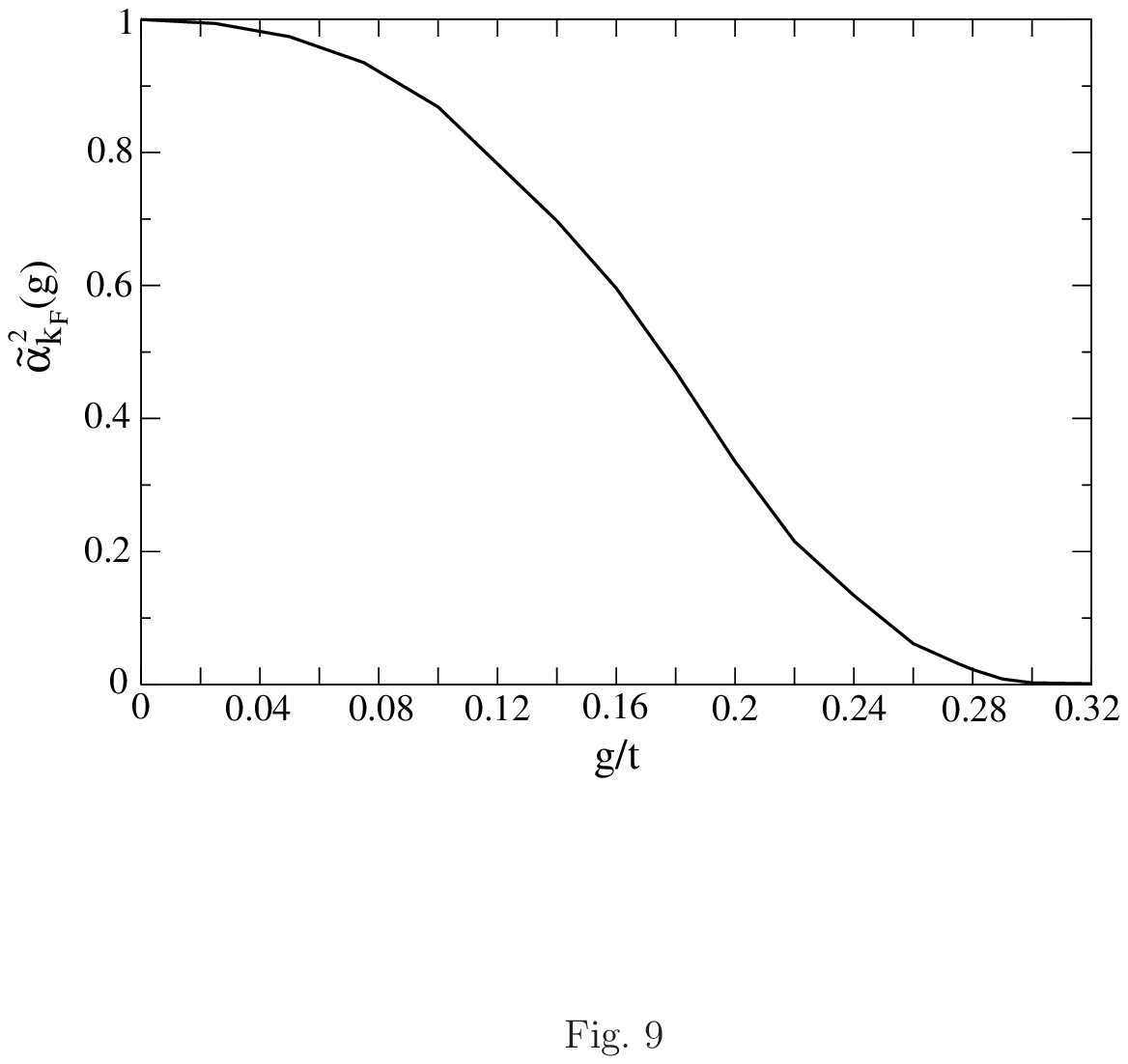}
  }
\end{center}

\end{document}